\pdfoutput=1
\documentclass[a4paper,12pt]{article}

\usepackage{amsfonts}
\usepackage{mathrsfs}
\usepackage{amsmath}
\usepackage{amssymb}
\usepackage{framed}

\usepackage{subfigure}
\usepackage{cancel}

\usepackage[medium]{titlesec}
\usepackage{bm}
\usepackage{cite}

\usepackage[normalem]{ulem}
\usepackage{extarrows}
\usepackage{slashed}
\usepackage{isodateo}
\usepackage{graphicx}
\usepackage[dvipsnames]{xcolor}
\usepackage[bookmarksnumbered=true,bookmarksopen=true]{hyperref}
 \hypersetup{colorlinks,%
 linkcolor=NavyBlue, %
             citecolor=PineGreen, %
             urlcolor=PineGreen}
\usepackage[hmargin=.7in,vmargin=1.1in]{geometry}
\usepackage{indentfirst}
\usepackage{booktabs}
\usepackage{placeins}
\usepackage{ulem} 

\usepackage{bbm}

\newcommand{\FR}[2]{\displaystyle\frac{\,{#1}\,}{#2}}

\newcommand{\n}{\nonumber}

\graphicspath{{fig/}}

\def\bge{\begin{equation}}
\def\ede{\end{equation}}
\def\bga{\begin{aligned}}
\def\eda{\end{aligned}}
\def\bgb{\begin{bmatrix}}
\def\edb{\end{bmatrix}}
\def\bgp{\begin{pmatrix}}
\def\edp{\end{pmatrix}}
\def\bgm{\begin{matrix}}
\def\edm{\end{matrix}}
\def\bgs{\begin{subequations}}
\def\eds{\end{subequations}}

\def\D{{\mathrm{d}}}

\def\pd{\partial}
\def\ld{{\mathscr{L}}}

\def\to{\rightarrow}

\def\I{\mathrm{i}}
\def\E{\mathrm{e}}

\def\lam{\lambda}

\def\si{\sigma}

\def\aa{\mathsf{a}}
\def\bb{\mathsf{b}}

\def\dd{\mathsf{d}}

\def\Re{\mathrm{Re}\,}

\def\2F1{{}_2\mathrm{F}_1}
\def\3F2{{}_3\mathrm{F}_2}

\usepackage{mdframed}

\newmdenv[skipabove=0pt,%
          skipbelow=5pt,%
          leftmargin=0pt,%
          rightmargin=0pt,%
          innertopmargin=-5pt,%
          innerbottommargin=7pt,%
          innerleftmargin=2pt,%
          innerrightmargin=2pt,%
          splittopskip=0pt,%
          splitbottomskip=0pt,%
          linewidth=0pt,%
          nobreak=true]%
          {keyeqn2}

\newmdenv[backgroundcolor=gray!15,%
          skipabove=0pt,%
          skipbelow=5pt,%
          leftmargin=0pt,%
          rightmargin=0pt,%
          innertopmargin=-5pt,%
          innerbottommargin=7pt,%
          innerleftmargin=2pt,%
          innerrightmargin=2pt,%
          splittopskip=0pt,%
          splitbottomskip=0pt,%
          linewidth=0pt,%
          nobreak=true]%
          {keyeqn}

\usepackage{titlesec}          
\titleformat{\section}
{\normalfont\fontsize{15}{20}\bfseries}{\thesection}{1em}{}

\newcommand{\wt}[1]{\mkern 2mu \widetilde{\mkern -2mu #1 \mkern -2mu}\mkern 2mu}
\newcommand{\wh}[1]{\mkern 2mu \widehat{\mkern-2mu#1\mkern-2mu}\mkern 2mu}

\newcommand{\fnemail}[1]{\footnote{Email: \href{mailto:#1}{\nolinkurl{#1}}}}

\title{\Large\textbf{New and Enhanced Cosmological Collider Shapes from Light--Heavy Field Mixing\\[2mm]
}}

\author{
Zhehan Qin,$^{\,a,b\,}$\fnemail{zq242@cam.ac.uk}~~~~
Jiayi Wu,$^{\,b\,}$\fnemail{wu-jy22@mails.tsinghua.edu.cn}~~~~
and~~~~Zhong-Zhi Xianyu$^{b,c\,}$\fnemail{zxianyu@tsinghua.edu.cn}
\\[5mm]
$^a\,$\normalsize{\emph{Department of Applied Mathematics and Theoretical Physics, University of Cambridge,}}\\\normalsize{\emph{Wilberforce Road, Cambridge, CB3 0WA, UK}}\\
$^b\,$\normalsize{\emph{Department of Physics, Tsinghua University, Beijing 100084, China} }\\
$^c\,$\normalsize{\emph{Peng Huanwu Center for Fundamental Theory, Hefei, Anhui 230026, China }}
}
\date{}

\allowdisplaybreaks[3]
\begin{document}
\maketitle

\begin{abstract}
We study the cosmological collider shapes generated by mixing a light scalar with a heavy scalar during inflation. The tree-level $s$-channel trispectrum generated by a boost-breaking two-point mixing between the two fields yields a new oscillatory shape in the leading mixed contribution, in which a power-law scaling factor from the light field multiplies the logarithmic oscillation from the principal field, giving a parametric enhancement in the squeezed limit. We also study the one-loop bispectrum generated by a bubble with one complementary and one principal internal line. Unlike a heavy-particle-only loop, whose nonanalytic signal is always subleading to its analytic background, the signal of the mixed loop can parametrically dominate the background in the squeezed limit for a range of light-field masses. Our results identify two concrete field-mixing mechanisms that enrich the shapes of cosmological collider signals and provide new phenomenological channels for studying flavor mixings at the cosmological collider.

\end{abstract}

\newpage
\tableofcontents
\clearpage

\section{Introduction}\label{introduction}

Cosmic inflation offers a distinctive window to study particle physics at extremely high energies~\cite{Achucarro:2022qrl}. The observed inflationary power spectrum suggests a (nearly) constant Hubble expansion parameter $H$ at least for a period of time during inflation, which makes de Sitter (dS) spacetime a good approximation for an inflationary universe. Current limits on the primordial tensor-to-scalar ratio suggest that $H$ is probably no higher than $10^{14}\,\mathrm{GeV}$~\cite{Planck:2018jri,BICEP:2021xfz}, while an inflation scenario close to this scale is preferable for a more natural realization of the inflation potential.

On a rapidly expanding background, quantum fields are stretched to super-horizon scales and left behind as classical perturbations. In particular, there may be pair productions directly from the dS vacuum of a field with mass comparable to $H$ and it subsequently decays into the inflaton, whose background value drives inflation. Together with gravity, this process leaves a distinctive non-Gaussian imprint on the correlation functions of the primordial fluctuations~\cite{Maldacena:2002vr}. Much as a terrestrial collider experiment reconstructs the physics from the debris of a collision, such correlators let us reconstruct the particle content and couplings present during inflation from the statistics of the fluctuations that survive to be observed today. This program is known as cosmological collider (CC) physics \cite{Chen:2009we,Chen:2009zp,Baumann:2011nk,Chen:2012ge,Pi:2012gf,Noumi:2012vr,Gong:2013sma,Arkani-Hamed:2015bza}. Thus, just like what terrestrial experiments do, CC takes the inflationary universe as the laboratory to probe high-energy physics.

The observable of CC is the cosmological correlator $\langle\varphi(\bm x_1)\cdots\varphi(\bm x_n)\rangle$, where $\varphi$ is the fluctuation of the inflaton evaluated on the future boundary of the dS inflationary patch. The exchange of a massive heavy scalar $\Sigma$ during inflation can imprint a logarithmically oscillating signal on such correlators in the squeezed limit, where the momentum $k_s$ carried by the exchanged particle is taken much softer than the external momenta, namely $k_{12}\equiv k_1+k_2$ (see Fig.~\ref{FigSingleExchange}): the signal takes the schematic form $A\cos[\widetilde\nu\log(k_s/k_{12})+\vartheta_0]$ , with the amplitude $A$ and the phase $\vartheta_0$ depending on details of the interaction. This oscillatory signal plays a dual role: Its frequency $\widetilde\nu$ from this oscillation directly reveals the exchanged particle's mass, while its logarithmic dependence on the momentum ratio is a direct consequence of exponential expansion~\cite{Chen:2014joa,Chen:2015lza,Chen:2018cgg,Quintin:2024boj}. For this reason, such a CC signal is also known as a primordial clock signal \cite{Chen:2011zf,Chen:2015lza,Arkani-Hamed:2015bza,Lee:2016vti}. For the above signal to be oscillatory, we generally need the heavy scalar $\Sigma$ to have mass $m_\Sigma>3H/2$ and thus mass parameter $\widetilde\nu=\sqrt{m_\Sigma^2/H^2-9/4}>0$, which is called a principal-series scalar in the terminology of dS representations \cite{Basile:2016aen,Sun:2021thf}. On the other hand, the scalar mass $0<m<3H/2$ corresponds to complementary series, which we dub ``light'' in this work. Such a scalar has a mass parameter $\nu=\sqrt{9/4-m^2/H^2}$ with $0<\nu<3/2$. The exchange of a complementary-series scalar imprints a qualitatively different signal on the same correlators: a nonoscillatory power law in the squeezed limit, $\propto (k_{s}/k_{12})^{-\nu}$ \cite{Chen:2009we,Chen:2009zp,Baumann:2011nk}.

Signals from both heavy and light scalars could be searched for from existing and future cosmological data such as the cosmic microwave background~\cite{Planck:2019kim,Sohn:2024xzd,Philcox:2024jpd,Philcox:2025bvj,Philcox:2025wts,Suman:2025tpv,Kumar:2026ogn,Kumar:2026dih}, large-scale structure survey~\cite{Ferraro:2022cmj,Meerburg:2016zdz,Cabass:2024wob,Green:2023uyz}, and more futuristic 21cm tomography~\cite{Munoz:2015eqa,Liu:2022iyy}. However, the two types of signals have different merits and drawbacks. For heavy scalars, the signal can reveal rich information through its oscillatory wiggles, but its amplitude typically carries a Boltzmann factor $\E^{-\pi\widetilde\nu}$~\cite{Chen:2012ge,Arkani-Hamed:2015bza,Lee:2016vti,Wang:2019gbi,Qin:2025xct}, reflecting the difficulty of producing a principal-series particle with $m\gg H$ from the dS vacuum. On the contrary, the power-law signal from a light scalar is controlled by its mass parameter $\nu$ via $(k_{s}/k_{12})^{-\nu}$~\cite{Chen:2009we,Chen:2009zp}. It carries no exponentially small Boltzmann factor and dilutes more slowly as the squeezed limit is approached, but is relatively featureless. For these reasons, the heavy oscillatory signal remains a unique and distinctive probe of the properties of the exchanged particle and the interaction, as has been extensively studied in recent years \cite{Chen:2016nrs,Chen:2016uwp,Chen:2016hrz,Lee:2016vti,An:2017hlx,Iyer:2017qzw,Kumar:2017ecc,Tong:2018tqf,Chen:2018sce,Chen:2018xck,Wu:2018lmx,Saito:2018omt,Li:2019ves,Lu:2019tjj,Liu:2019fag,Hook:2019zxa,Hook:2019vcn,Kumar:2018jxz,Kumar:2019ebj,Wang:2019gbi,Wang:2020uic,Li:2020xwr,Wang:2020ioa,Aoki:2020zbj,Bodas:2020yho,Lu:2021gso,Sou:2021juh,Lu:2021wxu,Pinol:2021aun,Cui:2021iie,Tong:2022cdz,Reece:2022soh,Chen:2022vzh,Niu:2022quw,Chen:2023txq,Chakraborty:2023qbp,Tong:2023krn,Jazayeri:2023xcj,Jazayeri:2023kji,Aoki:2023tjm,McCulloch:2024hiz,Craig:2024qgy,Pajer:2024ckd,Wang:2025qww,Bodas:2025vpb,Chen:2026khu,Ahmed:2026qtc,Aoki:2026nqr}, but the complementary-series signal's immunity to Boltzmann suppression and slower dilution make it also a valuable target in its own right~\cite{Green:2023uyz,Chakraborty:2023qbp,Goldstein:2024bky}.

The signals described above arise from the simplest case of a single exchanged field, with their qualitative character fixed by which representation the field belongs to. Existing studies of CC models mostly consider correlators of single-type exchange, either a light field \cite{Chen:2009we,Chen:2009zp,Baumann:2011nk,Goldstein:2024bky,Chakraborty:2025myb} or a heavy field \cite{Chen:2012ge,Noumi:2012vr,Chen:2015lza,Qin:2022fbv,Xianyu:2023ytd}. Scenarios with several additional scalars with perturbative or nonperturbative mixings have also been explored \cite{Emami:2013lma,McAneny:2019epy,Aoki:2020zbj,Pinol:2021aun,McAneny:2019epy,Werth:2026dcj,Green:2026yev,Philcox:2026tjj}. In this work, we explore an interesting possibility that the mixing between a light and a heavy field can combine the advantages of the two types of signal, retaining the oscillatory wiggles of a heavy signal and the squeezed-limit enhancement of a light signal.

There are two obvious ways to mix a heavy scalar $\si$ of mass $m_\sigma>3H/2$ with a light scalar $\pi$ with mass $0<m_\pi<3H/2$ within a bulk two-point function: either through a tree-level bilinear mixing or through a bubble loop; see Figs.~\ref{FigTree2pt} and \ref{FigLoopBispectrumDoubleFields}. We consider both situations in this work and reveal qualitatively new features in each of the resulting signals.

For the tree-level bilinear mixing of two scalars, we consider its contribution to the inflaton trispectrum. The mixing is nontrivial only when it breaks dS boosts. Since the dS boosts are broken by the rolling inflaton background, incorporating boost-breaking couplings, including bilinear mixings, into CC model building is a very natural option \cite{Cheung:2007st,Chen:2009we, Chen:2009zp, Baumann:2011nk, Noumi:2012vr,Lee:2016vti,An:2017hlx, Iyer:2017qzw,Chen:2018xck,McAneny:2019epy,Wang:2019gbi,Bodas:2020yho,Pinol:2021aun,Qin:2022fbv,Wang:2022eop,Tong:2022cdz,Pimentel:2022fsc,Jazayeri:2022kjy,Jazayeri:2023kji,Jazayeri:2023xcj,Aoki:2023tjm,Qin:2025xct,Xianyu:2025lbk,Werth:2026dcj,Green:2026yev}. In this work, we show that the boost-breaking  mixing $a^3\pi'\sigma$ introduces a genuinely new oscillatory shape, in which a complementary-series power law in one momentum ratio multiplies a principal-series logarithmic oscillation in the other.
This mixed shape is parametrically enhanced, relative to the corresponding principal-series-only oscillatory signal, by the weaker soft dilution of the complementary-series leg, though it remains subleading to the (nonoscillatory) leading complementary-series contribution to the same trispectrum.

In many models, the leading CC signal arises at one loop~\cite{Chen:2016uwp,Chen:2016hrz,Chen:2018xck,Lu:2019tjj,Hook:2019zxa,Kumar:2018jxz,Wang:2019gbi,Wang:2020ioa,Lu:2021gso,Cui:2021iie,Tong:2022cdz,Qin:2024gtr,Qin:2026yuo,Aoki:2026vbc,You:2026xoq}. For the one-loop bispectrum generated by a $\sigma$--$\pi$ bubble, the field mixing instead changes the signal--background hierarchy itself: the nonanalytic signal of an all-principal-series loop (see Fig.~\ref{FigLoopBispectrumSingleField}) always scales as $(k_3/k_{12})^4$ in the squeezed limit $k_3\to 0$ and is subleading to its $(k_3/k_{12})^3$ analytic background~\cite{Xianyu:2022jwk,Liu:2024xyi,Liu:2026jzn}, but with one complementary-series line the signal scales as $(k_3/k_{12})^{4-\nu_\pi}$ where $\nu_\pi=\sqrt{9/4-m_\pi^2/H^2}$ is the mass parameter of the light field $\pi$. Therefore, for $\nu_\pi>1$, it is going to parametrically dominate the background in the squeezed limit. This is a hierarchy that an principal-series-only loop cannot achieve.

We compute both examples using Schwinger--Keldysh (SK) perturbation theory in dS space \cite{Schwinger:1960qe,Feynman:1963fq,Keldysh:1964ud, Weinberg:2005vy,Chen:2017ryl}. For the tree diagram we use the partial Mellin--Barnes (PMB) representation of \cite{Qin:2022lva,Qin:2022fbv}, adapted from the Mellin-Barnes (MB) technique developed in \cite{Sleight:2019mgd,Sleight:2019hfp}, which converts the Hankel-function mode functions into Mellin powers and reduces the vertex-time integrals to an algebraic problem. For the loop diagram we use the spectral (K\"all\'en--Lehmann) decomposition of \cite{Marolf:2010zp,Hogervorst:2021uvp,Loparco:2023rug,Xianyu:2022jwk,Zhang:2025nzd,Grafe:2026qsm}, which expresses the bubble as a one-dimensional superposition of tree-level exchanges.

The paper is organized as follows. Sec.~\ref{Preliminaries} reviews the single-species exchange results used as a baseline throughout. Sec.~\ref{Tree} introduces the two-point mixing mechanism and presents the tree-level mixed trispectrum calculation. Sec.~\ref{Loop} develops the mixed-field loop mechanism and the corresponding bispectrum calculation. We conclude in Sec.~\ref{Conclusion}.

There are four appendices following the main text to provide all technical details. App.~\ref{Conventions} collects conventions of special functions and useful formulae. Detailed calculations of tree graphs mediated by a single field and by two mixing fields are given in App.~\ref{AppSingleExchange} and \ref{CalTree}, respectively. Finally, bubble diagrams composing of two general scalars are calculated in App.~\ref{CalLoop}.

\paragraph{Notations and conventions} Throughout this work, we use the mostly-plus spacetime metric, $\D s^{2}=a^{2}(\tau)\left(-\D\tau^{2}+\D\bm{x}^{2}\right)$, with $\tau\in(-\infty,0)$ the conformal time and $a(\tau)\equiv -1/(H\tau)$ the scale factor. We set the Hubble parameter $H\equiv 1$ throughout.

For brevity, we use contracted subscripts to denote sums, e.g. $s_{12}\equiv s_{1}+s_{2}$, $n_{1234}\equiv n_{1}+n_{2}+n_{3}+n_{4}$, and $k_{12}\equiv k_{1}+k_{2}$. Boldface letters such as $\bm k$ denote three-momenta, and the corresponding italic letters such as $k$ denote their magnitudes.

\section{Preliminaries: Single-Species Massive-Scalar Exchange}\label{Preliminaries}

Before turning to field mixing, let us first review some relevant results for correlators from a single-species exchange at both tree and loop levels, which will serve as building blocks for calculations involving multiple field species in Secs.~\ref{Tree} and~\ref{Loop}. Technical derivations can be found in the literature; see, e.g.,~\cite{Arkani-Hamed:2018kmz,Qin:2022fbv,Qin:2023ejc,Xianyu:2022jwk}. Here we quote the results needed below; the full expressions for the correlators are given in Apps.~\ref{AppSingleExchange} and~\ref{CalLoop}.

Consider a scalar field $\Sigma$ with mass $m$. The mode function \(u(k,\tau)\) then satisfies the momentum-space Klein--Gordon equation
\begin{equation}
\label{eq_KG}
(\tau^2\partial_\tau^2 -2\tau\partial_\tau + k^2\tau^2 + m^2 ) u(k,\tau) = 0.
\end{equation}
We are only interested in the principal and complementary series of scalar representations of \(\mathrm{SO}(4,1)\) \cite{Basile:2016aen,Sun:2021thf}, and the mode function is fixed after imposing the Bunch--Davies (BD) initial condition and canonical normalization, up to an irrelevant constant phase.  For a principal-series scalar,
\begin{equation}\label{PriModeF}
u(k,\tau) = \FR{\sqrt\pi}{2} \E^{-\pi\wt\nu/2}(-\tau)^{3/2}
\mathrm H_{\I\wt\nu}^{(1)}(-k\tau), \qquad \wt\nu \equiv \sqrt{m^2-\FR94}>0,
\end{equation}
whereas for a complementary-series scalar,
\begin{equation}\label{ComModeF}
u(k,\tau) = \FR{\sqrt\pi}{2} \E^{\I\pi\nu/2}(-\tau)^{3/2}
\mathrm H_{\nu}^{(1)}(-k\tau), \qquad \nu \equiv \sqrt{\FR94-m^2}, \qquad 0 < \nu < \FR32.
\end{equation}
In particular, the complementary mode function \eqref{ComModeF} follows from \eqref{PriModeF} by the continuation \(\widetilde\nu\to-\I\nu\).\footnote{The alternative continuation \(\widetilde\nu\to+\I\nu\) gives the same mode function by \(\mathrm H_{-\nu}^{(1)}(z)=\E^{\I\pi\nu}\mathrm H_{\nu}^{(1)}(z)\). We use \(\widetilde\nu\to-\I\nu\) throughout.} We therefore obtain complementary-series results by analytically continuing the corresponding principal-series expressions, which should be done for the full correlators.

The mode functions for the two series exhibit distinct late-time behavior as \(-k\tau\to0\):
\begin{align}\label{eq_uScaling}
u(k,\tau) \xrightarrow{-k\tau\to0}
\begin{cases}
(-\tau)^{3/2}(-k\tau)^{\pm\I\widetilde\nu}, & \text{principal series},\\
(-\tau)^{3/2}(-k\tau)^{-\nu}, & \text{complementary series}.
\end{cases}
\end{align}
Here the $\pm$ signs in the first line denote the two late-time branches of a principal-series scalar with their mass-dependent coefficients suppressed, and the second line shows only the dominant complementary-series branch.\footnote{For noninteger \(\nu\), the complementary-series mode also contains a subleading branch proportional to \((-k\tau)^{+\nu}\).} Therefore, at fixed \(k\), the principal-series mode oscillates logarithmically with an envelope proportional to \((-\tau)^{3/2}\), whereas the complementary-series mode decays more slowly as \((-\tau)^{3/2-\nu}\). These different late-time behaviors lead to different squeezed-limit scalings in exchange correlators, as illustrated below.

\subsection{Tree-level exchange}
\begin{figure}[t]
    \centering
    \includegraphics[width=0.48\textwidth]{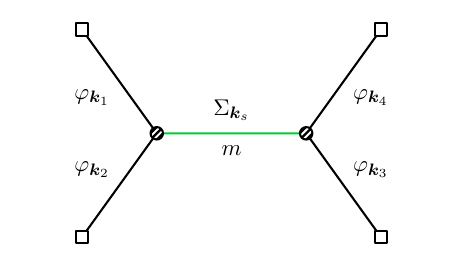}
    \hfill
    \includegraphics[width=0.48\textwidth]{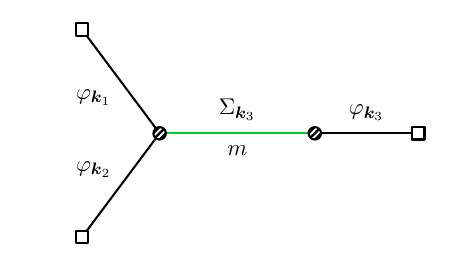}
    \caption{The $s$-channel single-exchange diagrams generating the four-point correlator $\mathcal T_s$ (left) and the three-point correlator $\mathcal B_s$ (right) through a massive scalar $\Sigma$.}
    \label{FigSingleExchange}
\end{figure}

\paragraph{Trispectrum} As a first example, we consider the inflaton four-point correlator \(\langle\varphi_{\bm k_1}\varphi_{\bm k_2} \varphi_{\bm k_3}\varphi_{\bm k_4}\rangle'\), where the prime means that the momentum-conserving factor \((2\pi)^3\delta^{(3)}(\bm k_1+\bm k_2+\bm k_3+\bm k_4)\) has been omitted.  We focus on its \(s\)-channel contribution \(\mathcal T_s\), generated by the exchange of a massive scalar \(\Sigma\) with mass \(m\) and momentum \(\bm k_s\equiv\bm k_1+\bm k_2\); see Fig.~\ref{FigSingleExchange}.

For definiteness, we choose a simple interaction that respects the shift symmetry of \(\varphi\),
\begin{equation}
\label{eq_LphiSigma}
   \ld \supset -\FR{1}2 a^2 \varphi'^2\Sigma,
\end{equation}
where we have omitted the overall dimensionful coupling. We compute the correlators using the SK formalism following the notation of~\cite{Chen:2017ryl}. With \(\mathsf a,\mathsf b=\pm\) labeling the two SK branches, the \(s\)-channel contribution reads
\begin{align}\label{eq_TsFeynman}
\mathcal T_s = \sum_{\mathsf a,\mathsf b=\pm}(-\mathsf a\mathsf b)\int_{-\infty}^0 \frac{\D \tau_1}{(-\tau_1)^2}\frac{\D \tau_2}{(-\tau_2)^2}\, K'_{\mathsf a}(k_1,\tau_1)K'_{\mathsf a}(k_2,\tau_1)K'_{\mathsf b}(k_3,\tau_2)K'_{\mathsf b}(k_4,\tau_2)D_{\mathsf a\mathsf b}(k_s;\tau_1,\tau_2),
\end{align}
where the massive bulk-to-bulk propagators are constructed from the mode function as
\begin{align}
&D_{-+}(k;\tau_1,\tau_2) = u(k,\tau_1)u^*(k,\tau_2),\qquad D_{+-}(k;\tau_1,\tau_2) = u^*(k,\tau_1)u(k,\tau_2),\\
&D_{\pm\pm}(k;\tau_1,\tau_2) = D_{\mp\pm}(k;\tau_1,\tau_2)\theta(\tau_1-\tau_2) + D_{\pm\mp}(k;\tau_1,\tau_2)\theta(\tau_2-\tau_1),
\end{align}
and the massless bulk-to-boundary propagators are
\begin{align}\label{eq_K}
K_\pm(k;\tau) = \frac{1\mp\I k\tau}{2k^3}\E^{\pm\I k\tau},\qquad K'_\pm(k;\tau) = \frac{\tau}{2k}\E^{\pm\I k\tau}.
\end{align}
Substituting \eqref{eq_K} into \eqref{eq_TsFeynman}, we express the trispectrum in terms of a dimensionless function $\mathcal I^{00}$ of two momentum ratios:
\begin{equation}
\label{eq_TvsI}
\mathcal T_s = \FR{1}{16k_1k_2k_3k_4k_s^5}\mathcal I^{00}(r_1,r_2),\qquad
r_1\equiv \FR{k_s}{k_{12}},\qquad r_2\equiv \FR{k_s}{k_{34}},
\end{equation}
where the general \emph{tree seed integral} $\mathcal I^{p_1p_2}$ introduced and computed in \cite{Qin:2022fbv}, with \(p_1,p_2\) denoting the two powers of time in the integrand, is defined by
\begin{align}\label{eq_Idef}
\mathcal I^{p_1p_2}\Big(\FR{k_s}{k_{12}},\FR{k_s}{k_{34}}\Big) = \sum_{\mathsf a,\mathsf b=\pm}(-\mathsf a\mathsf b)k_s^{5+p_{12}} \int_{-\infty}^0 \D\tau_1\D\tau_2\,  (-\tau_1)^{p_1}(-\tau_2)^{p_2}D_{\mathsf a\mathsf b}(k_s;\tau_1,\tau_2) \E^{\I\mathsf a k_{12}\tau_1+\I\mathsf b k_{34}\tau_2}.
\end{align}

The complete analytic result of the tree seed integral is collected in App.~\ref{AppSingleExchange}. Here we focus on the hierarchical squeezed limit \(k_s\ll k_{34}\ll k_{12}\), or equivalently \(r_1\ll r_2\ll1\). For $\Sigma$ being a principal-series scalar, the leading result is
\begin{align}
\label{eq_PriIs}
    \lim_{r_1\ll r_2\ll 1}\mathcal I^{00} =&~
    \FR{1+\I \sinh\pi\widetilde\nu}{2\pi}(r_1r_2)^{5/2}\bigg\{\Big(\FR{r_1r_2}{4}\Big)^{\I\widetilde\nu}\Gamma\Big[\FR52+\I\widetilde\nu,\FR52+\I\widetilde\nu,-\I\widetilde\nu,-\I\widetilde\nu\Big]\n\\
    &+\Big(\FR{r_1}{r_2}\Big)^{\I\widetilde\nu}\Gamma\Big[\FR52+\I\widetilde\nu,\FR52-\I\widetilde\nu,-\I\widetilde\nu,+\I\widetilde\nu\Big]\bigg\} + (\widetilde\nu\to-\widetilde\nu),
    \qquad \wt\nu>0,
\end{align}
whereas the leading result in the complementary-series case is\footnote{The expression below assumes generic masses. Results for special values of mass including \(\nu\in\{0,1/2,1\}\) can be obtained by carefully taking the limit.}
\begin{align}
\label{eq_ComIs}
    \lim_{r_1\ll r_2\ll 1}\mathcal I^{00} =&~
    \FR{1-\sin\pi\nu}{2\pi}(r_1r_2)^{5/2}\Big(\FR{r_1r_2}{4}\Big)^{-\nu}\Gamma\Big[\FR52-\nu,\FR52-\nu,+\nu,+\nu\Big]
    ,\qquad 0<\nu<\FR32.
\end{align}
The terms with characteristic mass-dependent momentum powers in \eqref{eq_PriIs} and \eqref{eq_ComIs} constitute the CC signals, dominating the full correlator in this squeezed limit. From the above expressions, we notice three useful differences:
\begin{itemize}
    \item \emph{Different shapes.} Apart from the common factor \((r_1r_2)^{5/2}\), the principal-series signal contains \((r_1r_2)^{\pm\I\widetilde\nu}\) and \((r_1/r_2)^{\pm\I\widetilde\nu}\), producing logarithmic oscillations with frequency \(\widetilde\nu\).\footnote{The terms proportional to \((r_1/r_2)^{\pm\I\widetilde\nu}\) and \((r_1r_2)^{\pm\I\widetilde\nu}\) are conventionally called the \emph{local} and \emph{nonlocal} signals, respectively, according to their analyticity in \(k_s\); see, e.g., \cite{Tong:2021wai}.} The leading complementary-series signal instead carries the nonoscillatory power \((r_1r_2)^{-\nu}\).\footnote{Other nonanalytic branches proportional to \((r_1/r_2)^{\pm\nu}\) or \((r_1r_2)^\nu\) are subleading for \(r_1,r_2<1\).}

    \item \emph{Boltzmann suppression.} In the large mass limit $m\to \infty$, or equivalently $\widetilde\nu\to\infty$ in the principal-series case, the oscillatory signal is
    \begin{equation}\label{eq_Boltzmann}
    \lim_{\widetilde\nu\to\infty}\lim_{r_1\ll r_2\ll 1}\mathcal I^{00} = -2\pi\widetilde{\nu}^3  \E^{-\pi\widetilde\nu}(r_1r_2)^{5/2}\bigg[
    \cos\Big(\widetilde\nu\log\FR{r_1r_2}{4}\Big) + \sin\Big(\widetilde\nu\log\FR{r_1}{r_2}\Big)\bigg].
    \end{equation}
    Thus, heavy principal-series exchange is exponentially suppressed by the Boltzmann factor $\E^{-\pi\widetilde\nu}$.

    \item \emph{Complementary-series amplification.} At fixed \(r_2\), the leading complementary-series signal scales as \(r_1^{5/2-\nu}\), whereas the principal-series signal has an envelope proportional to \(r_1^{5/2}\). The complementary-series signal is therefore parametrically enhanced in the squeezed limit.
\end{itemize}

\paragraph{Bispectrum} We next consider the tree-level bispectrum \(\mathcal B_s\equiv \langle\varphi_{\bm k_1}\varphi_{\bm k_2}\varphi_{\bm k_3}\rangle_s'\) generated by the same \(s\)-channel exchange in Fig.~\ref{FigSingleExchange}.  Here the exchanged momentum is \(k_s=k_3\).  The diagram requires an additional two-point mixing between \(\varphi\) and \(\Sigma\); for definiteness, we take
\begin{equation}
\ld \supset - a^3\varphi' \Sigma - \frac12 a^2\varphi'^2\Sigma,
\end{equation}
with which the $s$-channel bispectrum reads
\begin{align}\label{eq_BsTreeFeynman}
\mathcal B_s = \sum_{\mathsf a,\mathsf b=\pm}(-\mathsf a\mathsf b)\int_{-\infty}^0 \frac{\D \tau_1}{(-\tau_1)^2}\frac{\D \tau_2}{(-\tau_2)^3}\, K'_{\mathsf a}(k_1,\tau_1)K'_{\mathsf a}(k_2,\tau_1)K'_{\mathsf b}(k_3,\tau_2)D_{\mathsf a\mathsf b}(k_3;\tau_1,\tau_2).
\end{align}
Using \eqref{eq_K}, we can express the bispectrum in terms of a tree seed integral \eqref{eq_Idef} with $p_1=0$ and $p_2=-2$, evaluated at the folded limit $k_s/k_{34}\to 1$:
\begin{align}\label{eq:single-exchange-bispectrum-seed}
\mathcal B_s = -\frac{1}{8k_1k_2k_3^4}\mathcal I^{0,-2}(r,1), \qquad r \equiv \frac{k_3}{k_{12}}.
\end{align}
The finite folded limit of the seed is given in \eqref{eq_foldedI}; see more details in~\cite{Qin:2023ejc}. For a principal-series scalar, in the squeezed limit we have
\begin{align}\label{eq_tree_bis_sq_pri}
\lim_{r\to 0} \mathcal I^{0,-2}(r,1) =
-2^{-1/2-\I\wt\nu}\pi^{1/2} (\operatorname{sech}\pi\wt\nu+ \I \tanh \pi\wt\nu)r^{5/2+\I\wt\nu}
\Gamma\Big[ \frac{5}{2}+\I\wt\nu,-\I\wt\nu \Big] + (\wt\nu\to-\wt\nu),
\end{align}
whereas the leading complementary-series result is
\begin{align}\label{eq_tree_bis_sq_com}
\lim_{r\to 0} \mathcal I^{0,-2}(r,1) =
-2^{-1/2+\nu}\pi^{1/2} (\sec \pi\nu - \tan \pi\nu)r^{5/2-\nu}
\Gamma\Big[ \frac{5}{2}-\nu,\nu \Big].
\end{align}

\begin{figure}[t]
    \centering
    \includegraphics[width=0.48\textwidth]{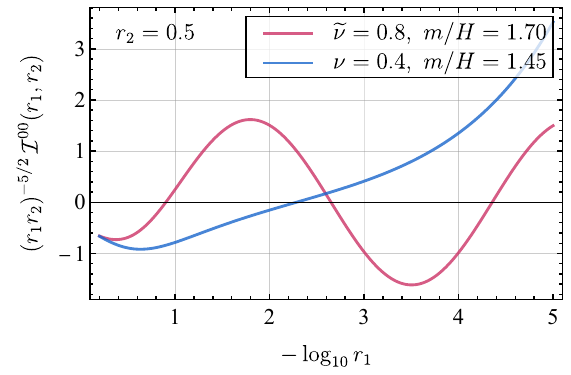}
    \hfill
    \includegraphics[width=0.48\textwidth]{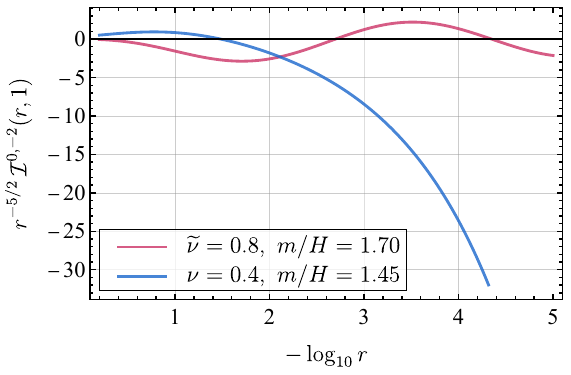}
    \caption{Cosmological correlators from single-species $s$-channel exchange, shown through the rescaled full tree seeds. Left: $(r_1r_2)^{-5/2}\mathcal I^{00}(r_1,r_2)$ as a function of $-\log_{10}r_1$, with $r_2=0.5$. Right: $r^{-5/2}\mathcal I^{0,-2}(r,1)$ as a function of $-\log_{10}r$. The seeds are related to the trispectrum and bispectrum by \eqref{eq_TvsI} and \eqref{eq:single-exchange-bispectrum-seed}, respectively. Red curves show principal-series exchange with $\widetilde\nu=0.8$, and blue curves show complementary-series exchange with $\nu=0.4$.}
    \label{fig_singletree}
\end{figure}

These expressions exhibit the same contrast as the trispectrum: CC signals dominate the squeezed limit, with logarithmic oscillations for principal-series exchange and weaker soft-momentum suppression for complementary-series exchange, as illustrated in Fig.~\ref{fig_singletree}.

\subsection{Loop-level bispectrum}
\begin{figure}[t]
    \centering
    \includegraphics[width=0.48\textwidth]{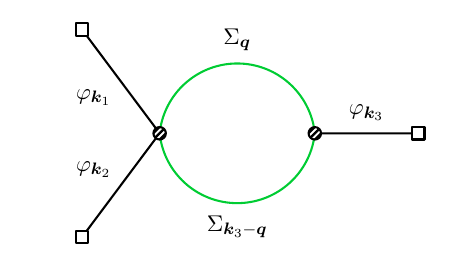}
    \caption{The $s$-channel bubble diagram that generates the bispectrum through the exchange of a pair of massive particles $\Sigma$.}
    \label{FigLoopBispectrumSingleField}
\end{figure}

Finally, we consider the \(s\)-channel bispectrum generated by a bubble of identical fields, as shown in Fig.~\ref{FigLoopBispectrumSingleField}. We choose the relevant interaction terms to be
\begin{equation}
    \label{eq:single-field-loop-interaction}
    \ld \supset -\FR{1}2a^3\varphi'\Sigma^2 -\FR{1}4a^2\varphi'^2\Sigma^2,
\end{equation}
with the overall couplings stripped. The Feynman rules then read
\begin{align}\label{eq_Bdef}
\mathcal B_s =&~ \FR12 \sum_{\mathsf a,\mathsf b=\pm}(-\mathsf a\mathsf b)\int_{-\infty}^0 \frac{\D \tau_1}{(-\tau_1)^2}\frac{\D \tau_2}{(-\tau_2)^3}\, K'_{\mathsf a}(k_1,\tau_1)K'_{\mathsf a}(k_2,\tau_1)K'_{\mathsf b}(k_3,\tau_2)\n\\
& \times \int \FR{\D^3\bm q}{(2\pi)^3}\,D_{\mathsf a\mathsf b}(|\bm q|;\tau_1,\tau_2)D_{\mathsf a\mathsf b}(|\bm k_3-\bm q|;\tau_1,\tau_2).
\end{align}
Here the prefactor \(1/2\) is the symmetry factor for two identical internal lines.  Substituting the external propagators gives
\begin{equation}\label{BinJ}
    \mathcal B_s = -\FR{1}{16k_1k_2k_3^4}
    \mathcal J_{\widetilde\nu,\widetilde\nu}(r),
    \qquad r\equiv \FR{k_3}{k_{12}},
\end{equation}
where, for later use, we define the unequal-mass \emph{loop seed integral} for the bispectrum by
\begin{align}\label{LoopSeedIntDef}
        \mathcal{J}_{\widetilde{\nu}_1,\wt\nu_2}\Big(\FR{k_3}{k_{12}}\Big)
        =&~
        \sum_{\mathsf a,\mathsf b=\pm}(-\mathsf a\mathsf b)\,k_{3}^3
        \int_{-\infty}^{0}\D\tau_{1}\FR{\D\tau_{2}}{(-\tau_2)^2}\,
        \E^{\I\mathsf{a}k_{12}\tau_{1}+\I\mathsf{b}k_{3}\tau_{2}}\n\\
&\times        \int \FR{\D^3\bm q}{(2\pi)^3}\,D^{(\wt\nu_1)}_{\mathsf a\mathsf b}(|\bm q|;\tau_1,\tau_2)D^{(\wt\nu_2)}_{\mathsf a\mathsf b}(|\bm k_3-\bm q|;\tau_1,\tau_2).
\end{align}
The mass labels on \(D_{\mathsf a\mathsf b}\), and correspondingly on the seed, are shown explicitly when different internal masses must be distinguished; otherwise they are suppressed.

\begin{figure}[t]
    \centering
    \includegraphics[width=0.48\textwidth]{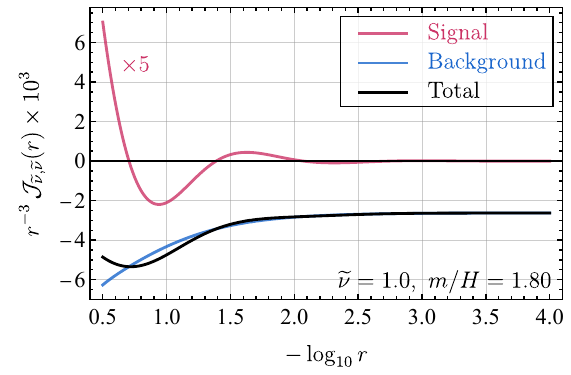}
    \hfill
    \includegraphics[width=0.48\textwidth]{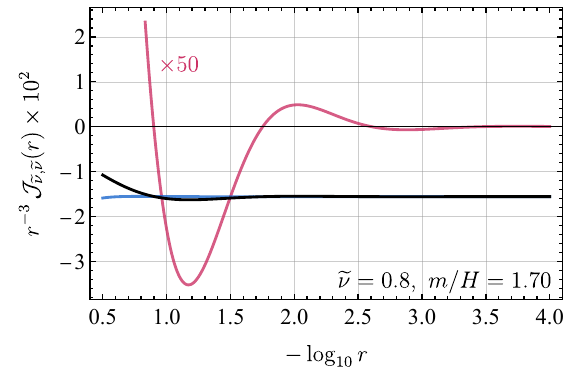}
    \\
    \includegraphics[width=0.48\textwidth]{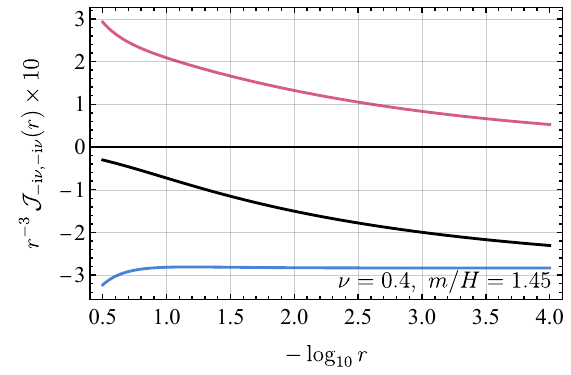}
    \hfill
    \includegraphics[width=0.48\textwidth]{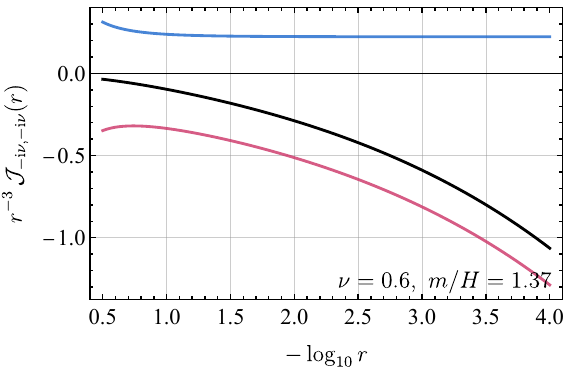}
    \caption{Rescaled single-species loop seeds \(r^{-3}\mathcal J(r)\)
    as functions of \(-\log_{10}r\).  The upper panels have
    \(\widetilde\nu=1.0\) and \(0.8\), and the lower panels have
    \(\nu=0.4\) and \(0.6\), respectively.  Red and blue denote the
    signal and renormalized background obtained from the complete seed
    \eqref{eq:loop-seed-degenerate} in App.~\ref{CalLoop},
    with \(\mu_R=H=1\); black shows their sum.
    Overall vertical factors of \(10^3\), \(10^2\), \(10\), and \(1\)
    are applied from the upper left to the lower right.
    The signal curves in the upper panels are additionally multiplied
    by \(5\) and \(50\), respectively, for visibility.}
    \label{fig_singleloop}
\end{figure}

We evaluate the loop seed using spectral decomposition \cite{Marolf:2010zp,Loparco:2023rug}, which expresses the bubble as a superposition of tree exchanges. Technical details, including the renormalization prescription, are given in App.~\ref{CalLoop}; more details can be found in \cite{Xianyu:2022jwk}. In the squeezed limit, the analytic background begins at \(r^3\) with a scheme-dependent coefficient, whereas the nonanalytic signal is UV finite and scheme independent. For a principal-series scalar, the leading signal is
\begin{align}\label{eq_J_Pri_sq}
\left.\lim_{r\to 0}\mathcal J_{\widetilde\nu,\widetilde\nu}(r)
\right|_{\mathrm{sig}}
={}& \left[\frac{-2\I\tanh \pi\wt\nu}{\pi^2}
 \Big(\frac r2\Big)^{4+2\I\wt\nu}
 \Gamma\Big[\frac32+\I\wt\nu,\frac52+\I\wt\nu,
 -\I\wt\nu,-\I\wt\nu\Big]
 + (\wt\nu\to-\wt\nu)\right].
\end{align}
For a complementary-series scalar, analytic continuation gives the leading signal
\begin{align}\label{eq_J_Com_sq}
\left.\lim_{r\to 0}\mathcal J_{-\I\nu,-\I\nu}(r)
\right|_{\mathrm{sig}}
={}& \frac{2\tan \pi\nu}{\pi^2}
 \Big(\frac r2\Big)^{4-2\nu}
 \Gamma\Big[\frac32-\nu,\frac52-\nu,\nu,\nu\Big].
\end{align}
Comparing these squeezed-limit scalings, we find that in contrast to the tree-level case, the principal-series signal \(r^{4\pm2\I\widetilde\nu}\) is subleading to the analytic \(r^3\) background whenever the latter has a nonzero renormalized coefficient. For \(\nu>1/2\), the complementary-series signal scales as \(r^{4-2\nu}\) and dominates the background.  Therefore, an all-principal loop supplies oscillations but no squeezed-limit dominance, whereas an all-complementary loop can be enhanced but is nonoscillatory, as shown in Fig.~\ref{fig_singleloop}.

\section{Tree-Level Mixed Trispectrum}\label{Tree}
As mentioned in the Introduction, we expect the slower late-time decay of a complementary-series field could enhance the oscillatory signal associated with a principal-series field.  In this section, we investigate this mechanism in a tree-level trispectrum generated by a two-point mixing between the two fields.  We refer to the resulting soft-scaling enhancement as an \emph{amplified oscillatory signal}.\footnote{Here and below, amplification is defined relative to the purely principal-series signal specified in each comparison, after stripping the common kinematic prefactor and holding the couplings and any common principal-series frequency fixed.  It does not by itself imply a larger complete correlator or signal-to-noise ratio.}

For the examples below, we denote the principal-series scalar by $\sigma$, with mass $m_\sigma>3/2$, and the complementary-series scalar by $\pi$, with mass $0<m_\pi<3/2$.  Consistent with \eqref{PriModeF} and \eqref{ComModeF}, their mass parameters are
\begin{equation}
    \widetilde{\nu}_\sigma = \sqrt{m_\sigma^2-\FR94},\qquad \nu_\pi = \sqrt{\FR94-m_\pi^2},
    \qquad \widetilde\nu_\sigma>0,\qquad 0<\nu_\pi<\FR32,
\end{equation}
and the free Lagrangian for the pair $(\sigma,\pi)$ is
\begin{equation}
\label{eq_freeL}
   \mathscr L_0[\sigma,\pi]= \sqrt{-g}\Big[ -\FR12 (\pd_\mu \sigma)^2-\FR12 m_\sigma^2\sigma^2 -\FR12 (\pd_\mu \pi)^2-\FR12 m_\pi^2\pi^2\Big].
\end{equation}

\subsection{Two-point mixing and interactions}
Consider first a covariant mass mixing,
\begin{equation}
\label{eq_Cov2pt}
 \ld \supset -\mu^2\sqrt{-g}\,\pi\sigma,
\end{equation}
where \(\mu^2\) is the mixing parameter.  This interaction does not produce a genuinely new mixed signal because it can be removed by a linear field redefinition.  Indeed, let \(\alpha\) be a real rotation parameter.  Applying
\begin{equation}
    \sigma\to \FR{\sigma+\alpha \pi}{\sqrt{1+\alpha^2}},\qquad \pi\to\FR{\pi-\alpha \sigma}{\sqrt{1+\alpha^2}},
\end{equation}
to \eqref{eq_freeL} gives
\begin{align}
    \mathscr L_0 \to& \sqrt{-g}\Big[ -\FR12 (\pd_\mu \sigma)^2 -\FR12  \FR{m_\sigma^2+\alpha^2 m_\pi^2}{1+\alpha^2}\sigma^2  -\FR12(\pd_\mu \pi)^2- \FR12\FR{m_\pi^2+\alpha^2 m_\sigma^2}{1+\alpha^2}\pi^2\n\\
    & - \FR{\alpha(m_\sigma^2-m_\pi^2)}{1+\alpha^2}\pi\sigma\Big].
\end{align}
The rotation therefore generates the off-diagonal term \eqref{eq_Cov2pt} with \(\mu^2=\alpha(m_\sigma^2-m_\pi^2)/(1+\alpha^2)\), while simultaneously changing the two diagonal mass entries as displayed above.  Conversely, any real symmetric quadratic mass matrix can be diagonalized by an orthogonal rotation.  An alternative invertible redefinition,
\begin{equation}
    \sigma \to \FR{1}{\sqrt{1+\alpha^2}}\sigma + \FR{\alpha m_\pi^2}{\sqrt{m_\sigma^4+\alpha^2m_\pi^4}}\pi,\qquad \pi \to \FR{m_\sigma^2}{\sqrt{m_\sigma^4+\alpha^2m_\pi^4}}\pi - \FR{\alpha}{\sqrt{1+\alpha^2}}\sigma,
\end{equation}
gives
\begin{align}
    \mathscr L_0 \to&~ \sqrt{-g}\Big[ -\FR12(\pd_\mu \sigma)^2 -\FR12 \FR{m_\sigma^2+\alpha^2 m_\pi^2}{1+\alpha^2}\sigma^2  -\FR12(\pd_\mu \pi)^2- \FR12\FR{(m_\sigma^2+\alpha^2 m_\pi^2)m_\sigma^2m_\pi^2}{m_\sigma^4+\alpha^2m_\pi^4}\pi^2\n\\
    & - \FR{\alpha(m_\pi^2-m_\sigma^2)}{\sqrt{1+\alpha^2}\sqrt{m_\sigma^4+\alpha^2m_\pi^4}}(\pd_\mu\pi)(\pd^\mu\sigma)\Big],
\end{align}
whose last term is a covariant derivative mixing \(\sqrt{-g}(\pd_\mu\pi)(\pd^\mu\sigma)\).

The field redefinitions above reflect a general constraint from dS symmetry.  A mixed two-point signal requires a nonanalytic contribution whose two late-time endpoints carry the scaling weights of distinct mass eigenmodes, whether they belong to the same or different series.  If the bulk two-point function preserves dS isometries, its boundary limit obeys the conformal Ward identities~\cite{Baumann:2019oyu}.  Translation, rotation, and dilation invariance allow a separated-point scalar correlator proportional to \(|\bm x|^{-\Delta_1-\Delta_2}\); special conformal invariance further requires \(\Delta_1=\Delta_2\).  The shadow pairing \(\Delta_1+\Delta_2=3\) instead gives a contact term, analytic in the exchanged momentum \cite{Qin:2023nhv}.  Since distinct scalar masses have neither equal nor shadow-paired weights, a mixed two-point signal can arise only if the bulk two-point function breaks dS isometries.

In inflation, this symmetry breaking is supplied by the background of the slow-roll inflaton \(\phi_0\) in \(\phi=\phi_0+\varphi\), where \(\varphi\) is the inflaton perturbation.  The nonzero \(\dot\phi_0\) selects a preferred time direction and breaks dS boost symmetry.

Motivated by this, we consider the simplest boost-breaking two-point mixing with a single conformal-time derivative acting on the complementary scalar $\pi$:
\begin{equation}
\label{eq_2ptmixing}
    \ld \supset - \lam_2 a^3 \pi'\sigma,
\end{equation}
which can arise by expanding the dimension-5 operator $\sqrt{-g}(\pd_\mu\phi\pd^\mu\pi)\sigma/\Lambda$ around the rolling background. So, the coupling $\lam_2$ is related to the cutoff scale $\Lambda$ by $\lam_2=\dot\phi_0/\Lambda$. Alternatively, one could place the temporal derivative on the principal scalar $\sigma$. After integration by parts, this differs from \eqref{eq_2ptmixing} only by a boundary term and a removable covariant mass mixing:
\begin{equation}
    a^3\sigma'\pi \sim - \sigma \pd_\tau(a^3\pi) = -3a^4\sigma\pi - a^3 \sigma\pi'.
\end{equation}
The first term on the right-hand side is the covariant mass mixing \eqref{eq_Cov2pt}.  For the nondegenerate masses considered here, removing this term changes the endpoint couplings at first order in the mixing, while mass shifts begin at second order.  We therefore focus on the mixing in \eqref{eq_2ptmixing}.

\begin{figure}[t]
    \centering
    \includegraphics[width=0.48\textwidth]{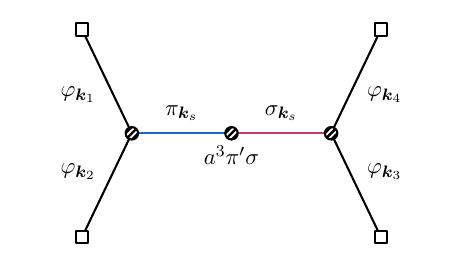}
    \hfill
    \includegraphics[width=0.48\textwidth]{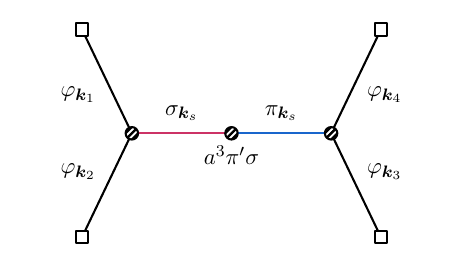}
    \caption{The two orientations of the $s$-channel trispectrum with a
    boost-breaking two-point mixing $a^3\pi'\sigma$.
    The diagrams are related by exchanging the endpoint energies
    $k_{12}\leftrightarrow k_{34}$; the derivative acts on $\pi$ in both
    orientations.}
    \label{FigTree2pt}
\end{figure}

Combining the mixing in \eqref{eq_2ptmixing} with endpoint interactions of the form \eqref{eq_LphiSigma}, and restoring the couplings, we take
\begin{equation}\label{Lagrangian2pt}
\Delta\ld
=-\lambda_2a^3\pi'\sigma
-\frac{\lambda_\sigma}{2}a^2\varphi'^2\sigma
-\frac{\lambda_\pi}{2}a^2\varphi'^2\pi.
\end{equation}
This interaction generates the two orientations contributing to the \(s\)-channel trispectrum, as shown in Fig.~\ref{FigTree2pt}. The same Lagrangian also generates the ordinary single-\(\sigma\) and single-\(\pi\) exchanges whose hierarchical limits are given in \eqref{eq_PriIs} and \eqref{eq_ComIs}.  Here and below, \(\mathcal T_s\) denotes only the contribution linear in \(\lambda_2\) shown in Fig.~\ref{FigTree2pt}.

We focus on the trispectrum because its two independent momentum ratios allow the scaling behaviors of the two massive fields to appear together in a single mixed shape.\begingroup\interfootnotelinepenalty=10000\footnote{With an appropriately modified endpoint interaction \cite{Qin:2023ejc}, tree-level bispectrum kinematics set \(r_2=k_s/k_{34}=1\) and \(r_1=k_s/k_{12}=k_3/k_{12}\equiv r\).  Although this point lies outside the hierarchical limit \(r_1\ll r_2\ll1\), kinematic counting suggests the schematic mappings \(r_1^{\pm\nu_\pi}r_2^{\pm\I\widetilde\nu_\sigma} \mapsto r^{\pm\nu_\pi}\) and \(r_1^{\pm\I\widetilde\nu_\sigma}r_2^{\pm\nu_\pi} \mapsto r^{\pm\I\widetilde\nu_\sigma}\).  The product dependence on two independent ratios is then lost, although both masses may still enter the coefficients.}\endgroup

\subsection{Trispectrum and seed integrals}
Following the SK Feynman rules, the contribution linear in \(\lambda_2\) is given by
\begin{align}\label{eq:tree-mixed-feynman}
\mathcal T_s
={}&
\lambda_2\lambda_\sigma\lambda_\pi
\sum_{\mathsf a,\mathsf b,\mathsf c=\pm}
(-\I\mathsf a)(-\I\mathsf b)(-\I\mathsf c)
\int_{-\infty}^{0}
\D\tau_L\D\tau_M\D\tau_R\,
a^2(\tau_L)a^3(\tau_M)a^2(\tau_R)
\nonumber\\
&\times
K'_{\mathsf a}(k_1,\tau_L)K'_{\mathsf a}(k_2,\tau_L)
K'_{\mathsf c}(k_3,\tau_R)K'_{\mathsf c}(k_4,\tau_R)
\nonumber\\
&\times
\bigg\{
\left[
\partial_{\tau_M}
D_{\mathsf a\mathsf b}^{(-\I\nu_\pi)}
(k_s;\tau_L,\tau_M)
\right]
D_{\mathsf b\mathsf c}^{(\widetilde\nu_\sigma)}
(k_s;\tau_M,\tau_R)
\nonumber\\
&\hspace{3.7cm}
+
D_{\mathsf a\mathsf b}^{(\widetilde\nu_\sigma)}
(k_s;\tau_L,\tau_M)
\left[
\partial_{\tau_M}
D_{\mathsf b\mathsf c}^{(-\I\nu_\pi)}
(k_s;\tau_M,\tau_R)
\right]
\bigg\}.
\end{align}
Substituting the external propagators \eqref{eq_K}, as in \eqref{eq_TvsI}, isolates a dimensionless seed \(\mathcal K\):
\begin{align}\label{T2K2ptMixed}
\mathcal T_s
={}&
\frac{\lambda_2\lambda_\sigma\lambda_\pi}
{16k_1k_2k_3k_4(k_{12}k_{34})^{5/2}}
\mathcal K(r_1,r_2),
\qquad
r_1=\frac{k_s}{k_{12}},
\qquad
r_2=\frac{k_s}{k_{34}},
\end{align}
where the seed \(\mathcal K\) is defined by the sum of the two oriented integrals,
\begin{align}
\mathcal K(r_1,r_2)
\equiv{}&\mathcal K_{\pi|\sigma}(r_1,r_2)
+\mathcal K_{\sigma|\pi}(r_1,r_2),
\nonumber\\
\mathcal K_{\pi|\sigma}(r_1,r_2)
\equiv{}&(k_{12}k_{34})^{5/2}
\sum_{\mathsf a,\mathsf b,\mathsf c=\pm}
(-\I\mathsf a)(-\I\mathsf b)(-\I\mathsf c)
\int_{-\infty}^{0}\D\tau_L\D\tau_M\D\tau_R\,
(-\tau_M)^{-3}
\nonumber\\
&\times\E^{\I\mathsf a k_{12}\tau_L+\I\mathsf c k_{34}\tau_R}
\left[\partial_{\tau_M}
D_{\mathsf a\mathsf b}^{(-\I\nu_\pi)}(k_s;\tau_L,\tau_M)\right]
D_{\mathsf b\mathsf c}^{(\widetilde\nu_\sigma)}(k_s;\tau_M,\tau_R),
\label{TreeSeed}\\
\mathcal K_{\sigma|\pi}(r_1,r_2)
\equiv{}&\mathcal K_{\pi|\sigma}(r_2,r_1).
\end{align}
The fields to the left and right of the vertical bar join the \(k_{12}\) and \(k_{34}\) endpoints, respectively, and the derivative acts on \(\pi\) in both orientations.  The last relation holds for the exact seeds; exchanging the arguments of a series restricted to \(r_1<r_2\) requires analytic continuation.  The general-mass seed definitions and their evaluation are given in App.~\ref{CalTree}, with the physical masses obtained by \(\widetilde\nu_1\to-\I\nu_\pi\) and \(\widetilde\nu_2\to\widetilde\nu_\sigma\).

\subsection{Components and hierarchical squeezed limit}
The seed \(\mathcal K\) can be decomposed into four components according to the analytic behavior of the correlator in the hierarchical squeezed limit \(r_1\ll r_2\ll1\):
\begin{align}
\mathcal K
={}&
\mathcal K_{\mathrm{mix}}
+
\mathcal K_{\mathrm{sca}}
+
\mathcal K_{\mathrm{osc}}
+
\mathcal K_{\mathrm{bg}}.
\label{eq:main-tree-component-decomposition}
\end{align}
Here \(\mathcal K_{\mathrm{mix}}\) contains nonanalytic powers of the momentum ratios associated with both fields.  The nonanalytic exponents in \(\mathcal K_{\mathrm{sca}}\) and \(\mathcal K_{\mathrm{osc}}\) are instead determined by the complementary- and principal-series weights, respectively.  The remaining component \(\mathcal K_{\mathrm{bg}}\) gives a contribution analytic in \(k_s^2\) and \(r_1/r_2\) once the momentum prefactor in \eqref{T2K2ptMixed} is restored.  Their exact definitions in terms of the factorized and nested contributions are given in \eqref{Kmix}--\eqref{Kbg}.

In particular, for generic masses, the leading scalings are 
\begin{equation}
\begin{aligned}
\lim_{r_1\ll r_2\ll1}\mathcal K_{\mathrm{mix}}(r_1,r_2)
&\sim r_1^{-\nu_\pi}r_2^{\pm\I\widetilde\nu_\sigma},
\\
\lim_{r_1\ll r_2\ll1}\mathcal K_{\mathrm{sca}}(r_1,r_2)
&\sim (r_1r_2)^{-\nu_\pi},
\\
\lim_{r_1\ll r_2\ll1}\mathcal K_{\mathrm{osc}}(r_1,r_2)
&\sim r_1^{\pm\I\widetilde\nu_\sigma}
       r_2^{\pm\I\widetilde\nu_\sigma},
\\
\lim_{r_1\ll r_2\ll1}\mathcal K_{\mathrm{bg}}(r_1,r_2)
&\sim \left(\frac{r_1}{r_2}\right)^{5/2}.
\end{aligned}
\label{eq:main-tree-component-scalings}
\end{equation}
Here \(\sim\) suppresses mass-dependent coefficients, and the two signs in \(\mathcal K_{\mathrm{osc}}\) are independent.  The mixed component combines the complementary field's power-law amplitude in \(r_1\) with the principal field's logarithmic oscillations in \(r_2\).  The displayed mixed scaling retains the dominant \(\pi|\sigma\) branch; the leading \(\sigma|\pi\) branch scales as \(r_1^{\pm\I\widetilde\nu_\sigma}r_2^{-\nu_\pi}\) and is suppressed by \((r_1/r_2)^{\nu_\pi}\).  The complementary-scaling component grows in both soft ratios and is leading in this limit, whereas the principal-series component has unit-modulus oscillatory powers.  The half-integer power in the background reflects the seed normalization.

\begin{figure}[t]
    \centering
    \includegraphics[width=0.48\textwidth]{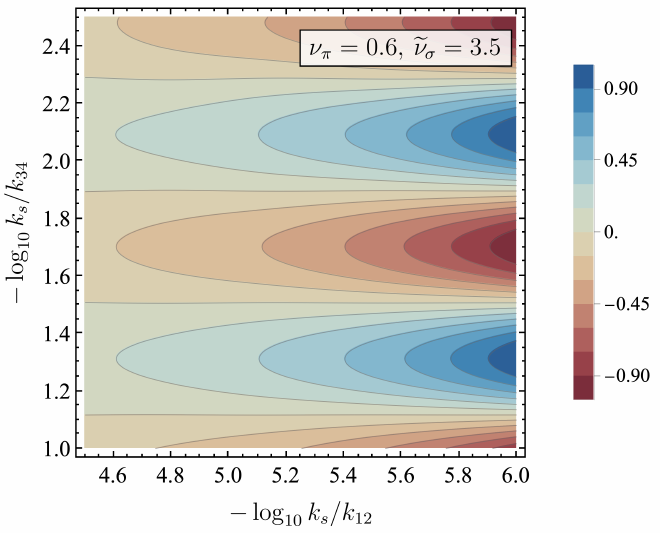}
    \hfill
    \includegraphics[width=0.48\textwidth]{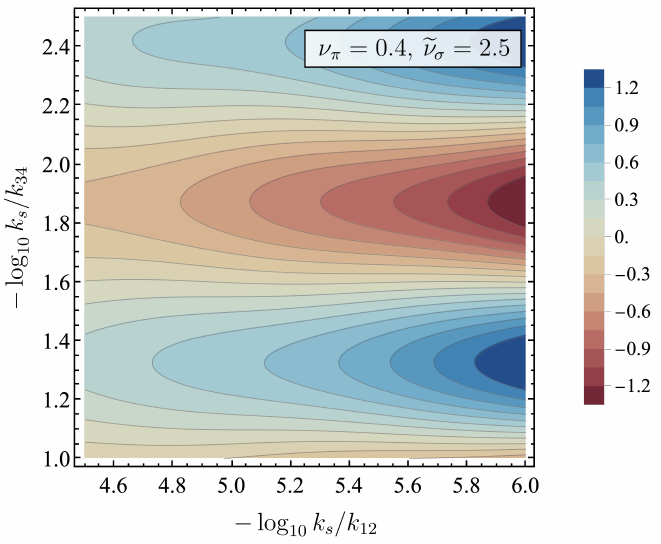}
    \caption{The full mixed component \(\mathcal K_{\mathrm{mix}}(r_1,r_2)\)
    in \eqref{eq:tree-mixed-exact}, including both orientations, both
    complementary-series branches, and all descendants.  The horizontal
    and vertical axes are \(-\log_{10}r_1\) and \(-\log_{10}r_2\),
    respectively.  The mass parameters are
    \((\nu_\pi,\widetilde\nu_\sigma)=(0.6,3.5)\) in the left panel and
    \((0.4,2.5)\) in the right panel.  The color scales give
    \(\mathcal K_{\mathrm{mix}}\) without additional rescaling.}
    \label{TreePheno}
\end{figure}

Resumming the fully factorized residue series gives the complete mixed component in closed form:
\begin{align}
\mathcal K_{\mathrm{mix}}(r_1,r_2)
={}&\sum_{\aa,\bb=\pm1}
\frac{
\sin\left[\frac\pi2\left(\frac52-\aa\nu_\pi\right)\right]
\sin\left[\frac\pi2\left(\frac52+\I\bb\widetilde\nu_\sigma\right)\right]
}{2\sin\left[\frac\pi2\left(\aa\nu_\pi+\I\bb\widetilde\nu_\sigma\right)\right]}\Gamma\left[
\aa\nu_\pi,\frac52-\aa\nu_\pi,
-\I\bb\widetilde\nu_\sigma,\frac52+\I\bb\widetilde\nu_\sigma
\right]
\nonumber\\
&\times\Bigg\{
 \left(\frac{r_1}2\right)^{-\aa\nu_\pi}\left(\frac{r_2}2\right)^{\I\bb\widetilde\nu_\sigma}
 {}_2\mathrm F_1\left[\begin{matrix}\frac54-\frac{\aa\nu_\pi}{2},
              \frac74-\frac{\aa\nu_\pi}{2}\\1-\aa\nu_\pi\end{matrix}\middle|r_1^2\right]
 {}_2\mathrm F_1\left[\begin{matrix}\frac54+\frac{\I\bb\widetilde\nu_\sigma}{2},
              \frac74+\frac{\I\bb\widetilde\nu_\sigma}{2}\\
              1+\I\bb\widetilde\nu_\sigma\end{matrix}\middle|r_2^2\right]\nonumber\\
 &+(r_1\leftrightarrow r_2)\Bigg\}.
\label{eq:tree-mixed-exact}
\end{align}
Here \({}_2\mathrm F_1\) is the ordinary Gauss hypergeometric function. The signs \(\aa\) and \(\bb\) label the two branches of \(\pi\) (complementary series) and \(\sigma\) (principal series), respectively. The replacement $(r_1\leftrightarrow r_2)$ acts on the entire product inside braces.  Equation~\eqref{eq:tree-mixed-exact} includes both orientations and all descendants for \(0<r_1,r_2<1\); its derivation is given in App.~\ref{CalTree}.  The full mixed component in \eqref{eq:tree-mixed-exact} is shown in Fig.~\ref{TreePheno}.

In the hierarchical squeezed limit, the hypergeometric functions approach unity and the dominant mixed branch gives
\begin{align}
\lim_{r_1\ll r_2\ll1}\mathcal K_{\mathrm{mix}}(r_1,r_2)
={}&\sum_{\bb=\pm1}
\frac{
\sin\left[\frac\pi2\left(\frac52-\nu_\pi\right)\right]
\sin\left[\frac\pi2\left(\frac52+\I\bb\widetilde\nu_\sigma\right)\right]
}{2\sin\left[\frac\pi2\left(\nu_\pi+\I\bb\widetilde\nu_\sigma\right)\right]}
\Gamma\left[
\nu_\pi,\frac52-\nu_\pi,
-\I\bb\widetilde\nu_\sigma,\frac52+\I\bb\widetilde\nu_\sigma
\right]
\nonumber\\
&\times
\left(\frac{r_1}{2}\right)^{-\nu_\pi}
\left(\frac{r_2}{2}\right)^{\I\bb\widetilde\nu_\sigma}.
\label{K120}
\end{align}

At fixed \(\nu_\pi\), taking the large mass limit \(\widetilde\nu_\sigma\gg1\) further simplifies \eqref{K120} to
\begin{align}
\lim_{r_1\ll r_2\ll1}
\mathcal K_{\mathrm{mix}}(r_1,r_2)
\simeq{}&
2\pi
\sin\left[\frac\pi2\left(\frac52-\nu_\pi\right)\right]
\Gamma\left[\nu_\pi,\frac52-\nu_\pi\right]
\widetilde\nu_\sigma^{3/2}
\E^{-\pi\widetilde\nu_\sigma}
\nonumber\\
&\times
\left(\frac{r_1}{2}\right)^{-\nu_\pi}
\cos\left[
\widetilde\nu_\sigma\log\left(\frac{r_2}{2}\right)
+\frac{\pi\nu_\pi}{2}
\right].
\label{K12Nu}
\end{align}

As shown in Fig.~\ref{TreePheno}, the dominant oscillations in the plotted hierarchy \(r_1\ll r_2\ll1\) run only along the \(r_2\) direction, while varying \(r_1\) mainly changes their power-law amplitude. The principal-series field sets the logarithmic oscillation frequency and contributes the single Boltzmann factor \(\E^{-\pi\widetilde\nu_\sigma}\).  The complementary-series field supplies the power-law amplitude \(r_1^{-\nu_\pi}\) and the phase shift \(\pi\nu_\pi/2\), without an additional Boltzmann suppression.

The fully factorized contribution also admits a simple on-shell interpretation: the boost-breaking insertion may be viewed as producing a correlated \(\sigma\)--\(\pi\) pair from the rolling background, with the two modes coupling to the external inflaton pairs at the endpoint vertices.  Their distinct late-time scaling weights generate the mixed momentum dependence.

After rewriting \eqref{eq_PriIs} with the same kinematic prefactor as \eqref{T2K2ptMixed}, the ordinary single-\(\sigma\) exchange contains only unit-modulus oscillatory powers.  The additional amplitude \(r_1^{-\nu_\pi}\) in \eqref{K120} therefore enhances the mixed signal in the soft scaling relative to the single-\(\sigma\) signal with the same frequency.  Both \(\mathcal K_{\mathrm{sca}}\) and the ordinary single-\(\pi\) exchange \eqref{eq_ComIs} instead carry \((r_1r_2)^{-\nu_\pi}\), so the mixed power-law amplitude is smaller by \(r_2^{\nu_\pi}\).  These comparisons concern momentum scaling; absolute ratios also depend on the couplings and mass-dependent coefficients. The remaining reflected branches and descendants are discussed in App.~\ref{CalTree}.

\section{Mixed-Field Loop Bispectrum}\label{Loop}

Unlike the tree-level bilinear mixing considered in Sec.~\ref{Tree}, a nontrivial mixing in a bubble diagram does not require dS boost breaking in the exchanged bulk two-point function, as long as the bubble couples to external lines in a boost-breaking way. For instance, the simplest exchange is described by the two-point function of the composite operator \(\pi\sigma\), which at leading order is simply the product of the two dS-invariant free propagators and allowed by dS isometry in the bulk.

\begin{figure}[t]
    \centering
    \includegraphics[width=0.48\textwidth]{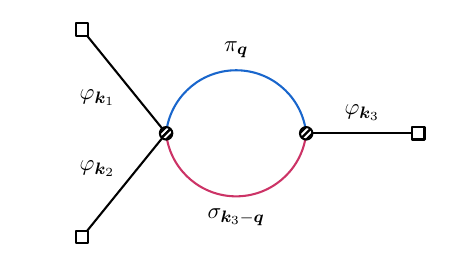}
    \caption{The Feynman diagram for a bubble bispectrum correlator, with a light and a heavy massive field.}
    \label{FigLoopBispectrumDoubleFields}
\end{figure}

However, a nonvanishing CC signal in the bispectrum on the future boundary still requires dS boost breaking effect \cite{Aoki:2026smw}, and we introduce this breaking through the couplings of \(\pi\) and \(\sigma\) to the external inflaton \(\varphi\). A simple choice is
\begin{equation}
\Delta\ld
=-\lambda_3a^3\varphi'\pi\sigma
-\frac{\lambda_4}{2}a^2\varphi'^2\pi\sigma.
\label{eq:mixed-loop-interaction}
\end{equation}
These interactions generate the bubble bispectrum shown in Fig.~\ref{FigLoopBispectrumDoubleFields}.

The SK Feynman rules give its \(s\)-channel contribution as
\begin{align}
\mathcal B_s
={}&
\lambda_3\lambda_4
\sum_{\mathsf a,\mathsf b=\pm}
(-\mathsf a\mathsf b)
\int_{-\infty}^{0}
\frac{\D\tau_1}{(-\tau_1)^2}
\frac{\D\tau_2}{(-\tau_2)^3}
K'_{\mathsf a}(k_1,\tau_1)
K'_{\mathsf a}(k_2,\tau_1)
K'_{\mathsf b}(k_3,\tau_2)
\nonumber\\
&\times
\int\frac{\D^3\bm q}{(2\pi)^3}\,
D_{\mathsf a\mathsf b}^{(-\I\nu_\pi)}
(|\bm q|;\tau_1,\tau_2)
D_{\mathsf a\mathsf b}^{(\widetilde\nu_\sigma)}
(|\bm k_3-\bm q|;\tau_1,\tau_2).
\label{eq:mixed-loop-feynman}
\end{align}
Because the two internal fields are distinct, there is no diagrammatic symmetry factor.  Substituting the external propagators and using the unequal-mass loop seed defined in \eqref{LoopSeedIntDef}, we obtain
\begin{equation}
\mathcal B_s
=
-\frac{\lambda_3\lambda_4}{8k_1k_2k_3^4}
\mathcal J_{-\I\nu_\pi,\widetilde\nu_\sigma}(r),
\qquad
r=\frac{k_3}{k_{12}}.
\label{eq:mixed-loop-bispectrum-seed}
\end{equation}
The full bispectrum follows by adding the two cyclic permutations.

The seed separates into a nonanalytic signal and a renormalized analytic background.  For \(0<r<1\), the complete signal obtained from \eqref{eq:loop-seed-unequal-mass} is
\begin{align}
\left.\mathcal J_{-\I\nu_\pi,\widetilde\nu_\sigma}(r)
\right|_{\mathrm{sig}}
={}&
\sum_{\aa,\bb=\pm1}
\frac{1-\cosh\left[\pi\left(
\bb\widetilde\nu_\sigma-\I\aa\nu_\pi\right)\right]}{16\pi^4}
\sum_{n=0}^{\infty}\frac{(-1)^n}{n!}\left(\frac{2r}{1+r}\right)^{4+2n-\aa\nu_\pi-\I\bb\widetilde\nu_\sigma}
\nonumber\\
&\times
\frac{\frac32+2n-\aa\nu_\pi-\I\bb\widetilde\nu_\sigma}
{\left(1+2n-\aa\nu_\pi-\I\bb\widetilde\nu_\sigma\right)
 \left(2+2n-\aa\nu_\pi-\I\bb\widetilde\nu_\sigma\right)}
\nonumber\\
&\times\Gamma\left[
\frac32+n,\frac32+n-\aa\nu_\pi,
\frac32+n-\I\bb\widetilde\nu_\sigma,
\frac32+n-\aa\nu_\pi-\I\bb\widetilde\nu_\sigma
\right]
\nonumber\\
&\times\Gamma\left[
-n+\aa\nu_\pi,-n+\I\bb\widetilde\nu_\sigma,
-n+\aa\nu_\pi+\I\bb\widetilde\nu_\sigma
\right]
\nonumber\\
&\times{}_2\mathcal F_1\left[\begin{matrix}
2+2n-\aa\nu_\pi-\I\bb\widetilde\nu_\sigma,
4+2n-\aa\nu_\pi-\I\bb\widetilde\nu_\sigma\\
4+4n-2\aa\nu_\pi-2\I\bb\widetilde\nu_\sigma
\end{matrix}\middle|\frac{2r}{1+r}\right].
\label{eq:mixed-loop-exact-signal}
\end{align}
Here \({}_2\mathcal F_1\) is the dressed hypergeometric function defined in \eqref{eq:dressed-hypergeometric}, and the sum over \(\aa,\bb\) includes all four mass branches. The derivation and the complete renormalized background are given in App.~\ref{CalLoop}.

In the squeezed limit, the \(n=0\) terms of \eqref{eq:mixed-loop-exact-signal} and the zero-argument value of the dressed hypergeometric function give the leading terms of the nonanalytic branches:
\begin{align}
\left.
\lim_{r\to0}
\mathcal J_{-\I\nu_\pi,\widetilde\nu_\sigma}(r)
\right|_{\mathrm{sig}}
={}&
\bigg\{
\frac{\I}{\pi^2}
\tanh\left[
\frac\pi2\left(\widetilde\nu_\sigma-\I\nu_\pi\right)
\right]
\left(3-\nu_\pi-\I\widetilde\nu_\sigma\right)
\nonumber\\
&\times
\Gamma\left[
\frac32-\nu_\pi,
\frac32-\I\widetilde\nu_\sigma,
\nu_\pi,
\I\widetilde\nu_\sigma
\right]
\left(\frac r2\right)^{
4-\nu_\pi-\I\widetilde\nu_\sigma}
\nonumber\\
&+
\left(\widetilde\nu_\sigma
\to-\widetilde\nu_\sigma\right)
\bigg\}
+
\left(\nu_\pi\to-\nu_\pi\right).
\label{eq:mixed-loop-squeezed-signal}
\end{align}
The last replacement acts on the preceding braces.  The \((\nu_\pi\to-\nu_\pi)\) pair is suppressed by \(r^{2\nu_\pi}\), so the leading signal has powers \(r^{4-\nu_\pi\pm\I\widetilde\nu_\sigma}\).  The renormalized analytic background is the continued \(r^3\) term in \eqref{eq:loop-seed-unequal-mass-squeezed}. Combining the leading signal and background gives the schematic bispectrum scaling
\begin{align}\label{eq:mixed-loop-scaling-preview}
    \lim_{r\to 0}\mathcal B_s
    \sim \frac{1}{k_1k_2k_3^4}
    \left(r^{4-\nu_\pi\pm\I\wt\nu_\sigma}+r^3\right).
\end{align}
Here the couplings and mass-dependent coefficients are omitted.

For a nonzero renormalized background coefficient, the signal-to-background envelope scales as \(r^{1-\nu_\pi}\).  The signal therefore dominates for \(1<\nu_\pi<3/2\), corresponding to \(0<m_\pi<\sqrt5/2\), and is subleading for \(0<\nu_\pi<1\).  At \(\nu_\pi=1\), the two envelopes have the same power, while the signal remains logarithmically oscillatory.

\begin{figure}[t]
    \centering
    \includegraphics[width=0.48\textwidth]{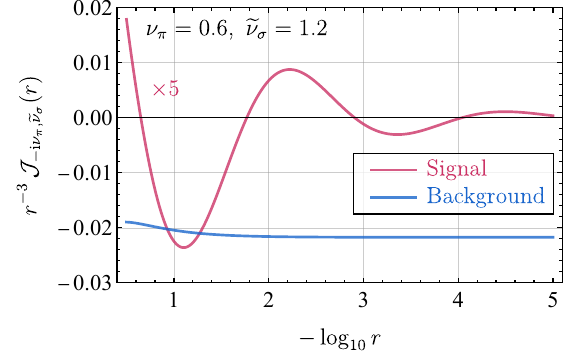}\hfill
    \includegraphics[width=0.48\textwidth]{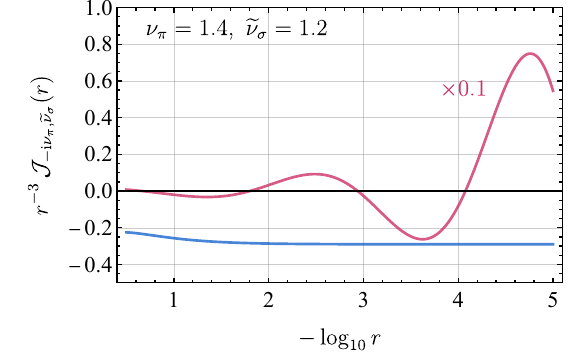}
    \caption{The signal (red) and renormalized background (blue) of the
    rescaled loop seed \(r^{-3}\mathcal J_{-\I\nu_\pi,\widetilde\nu_\sigma}(r)\),
    plotted against \(-\log_{10}r\).  The complete signal is given by
    \eqref{eq:mixed-loop-exact-signal}, and the complete renormalized
    background by the final two terms of \eqref{eq:loop-seed-unequal-mass}
    in App.~\ref{CalLoop}, with the physical continuation
    \eqref{eq:loop-physical-continuation} and \(\mu_R=H=1\).
    The mass parameters are
    \((\nu_\pi,\widetilde\nu_\sigma)=(0.6,1.2)\) in the left panel and
    \((1.4,1.2)\) in the right panel.  The signals in the left and right panels are
    multiplied by \(5\) and \(0.1\), respectively, for visibility.  The signal envelope decays
    in the left panel and grows in the right panel as \(r\to0\).}
    \label{LoopPheno}
\end{figure}

For comparison, an all-principal unequal-mass loop has a nonanalytic power-law amplitude proportional to \(r^4\) in \eqref{eq:loop-seed-unequal-mass-squeezed}; the complementary-series line therefore supplies the additional factor \(r^{-\nu_\pi}\) \cite{Xianyu:2022jwk}. This compares momentum scaling; absolute amplitudes also depend on the couplings and mass-dependent coefficients.  Using \eqref{GammaLimit}, for \(\widetilde\nu_\sigma\gg1\) the Gamma product in \eqref{eq:mixed-loop-squeezed-signal} carries one \(\E^{-\pi\widetilde\nu_\sigma}\) Boltzmann factor, with no additional Boltzmann suppression from the complementary-series line. Representative cases on the two sides of the threshold \(\nu_\pi=1\) are shown in Fig.~\ref{LoopPheno}.

\section{Conclusions and outlook}\label{Conclusion}
CC signals probe heavy particles present during
inflation, through their imprints on cosmological correlators with characteristic oscillatory momentum dependence.
Their observability, however, can be limited by Boltzmann suppression
and by their scaling relative to other contributions in the squeezed
limit. In this work, we have studied two mechanisms involving a
principal-series scalar and a complementary-series scalar: a
boost-breaking bilinear mixing that contributes to the tree-level
trispectrum, and a bubble containing one light and one heavy internal
line that generates a one-loop bispectrum. In both cases, the light
field enhances the squeezed-limit scaling of the oscillatory signal,
while the Boltzmann suppression associated with the heavy field remains.

At tree level, the bilinear mixing produces a distinctive trispectrum component that combines the scaling behaviors of the two fields in different momentum ratios. For generic masses, its dominant branch in the hierarchical squeezed limit \(r_1\ll r_2\ll1\) takes the schematic form \( \mathcal K_{\mathrm{mix}} \sim r_1^{-\nu_\pi} \cos(\widetilde\nu_\sigma\log r_2+\vartheta) \) with a mass-dependent phase $\vartheta$. The complementary-series field supplies the power-law envelope, while the principal-series field determines the logarithmic oscillation frequency. This mixed component is enhanced in its soft momentum scaling relative to the corresponding purely heavy-field signal, but remains subleading to the nonoscillatory complementary-series contribution in the same hierarchical limit.

At one loop, the light field can reverse the hierarchy between the oscillatory signal and the analytic EFT background. After factoring out the common kinematic prefactor, an all-principal-series bubble has a signal envelope proportional to \(r^4\) in the squeezed limit $r\to 0$, together with the background beginning at \(r^3\) \cite{Xianyu:2022jwk}, hence the bispectrum signal would be buried in the squeezed limit. With one internal line replaced by a light field, the oscillatory signal instead has an envelope proportional to \(r^{4-\nu_\pi}\). Consequently, for \(1<\nu_\pi<3/2\), the oscillatory signal parametrically dominates full correlator in the squeezed limit, as illustrated in Fig.~\ref{LoopPheno}. The mixed loop thus combines the logarithmic oscillations of a heavy field with a squeezed-limit dominance that the heavy bubble exchange cannot exhibit.

Taken together, these two examples show that the cosmological collider can probe the mixing between spectator fields of different masses. In both ways, a single oscillatory shape encodes the two fields at once: the oscillation frequency is set by the heavy mass through $\widetilde\nu_\sigma$, while the power-law envelope is controlled by the light mass through $\nu_\pi$. The mixed loop is of particular interest in this respect, since its oscillatory signal can parametrically dominate the analytic background in the squeezed limit for $1<\nu_\pi<3/2$. Such mixed signals therefore provide a new channel for studying flavor mixings at the cosmological collider.

Several theoretical extensions merit further investigation. Recent analytic and nonperturbative studies of mixing between curvature perturbations and massive fields motivate extending the present light--heavy system beyond perturbative mixing~\cite{An:2017hlx,Iyer:2017qzw,Huenupi:2026abj,Huenupi:2026aqc, Wang:2026lff,Pinol:2026xnl,Pinol:2026hey,Belrhali:2026uxn, Huenupi:2026jcv,Werth:2026dcj}.
It would be particularly interesting to study how strong mixing between the two massive fields modifies the scaling exponents and oscillatory shapes found in this work~\cite{McAneny:2019epy,Werth:2026dcj,Green:2026yev}. Extending the analysis to spinning fields could reveal mixed signals with additional angular structure \cite{Arkani-Hamed:2015bza,Lee:2016vti}. In addition, combining the light--heavy mixing with reduced sound speeds \cite{Pimentel:2022fsc,Jazayeri:2022kjy,Jazayeri:2023xcj} or chemical potentials \cite{Chen:2018xck,Wang:2019gbi,Bodas:2020yho,Qin:2022fbv} could improve the visibility of the mixed signal over a finite range of momentum configurations. It would also be valuable to realize such light--heavy mixings in more concrete particle model building beyond the Standard Model \cite{Chen:2016uwp,Chen:2016hrz,Kumar:2017ecc,Kumar:2018jxz, Lu:2019tjj,Hook:2019zxa,Wang:2020ioa}. Such realizations would relate the masses and mixing couplings to underlying model parameters, and dedicated templates for both the mixed bispectra and trispectra would enable searches in cosmological data~\cite{Philcox:2025bvj,Philcox:2025wts,Suman:2025vuf,Green:2026yev,Philcox:2026tjj,Philcox:2026njr}. We leave all these directions to future work.

\paragraph{Acknowledgments} We thank Xingang Chen for initial collaboration and Shuntaro Aoki, Xi Tong, Lian-Tao Wang, Denis Werth, Hongyu Zhang, Yisong Zhang, and Yuhang Zhu for useful discussions. 
This work is supported by NSFC under Grants No.\ 12275146, No.\ 12247103, No.\ 12521007, and the Dushi Program of Tsinghua University. Z.Q. is partly supported by the China Scholarship Council (CSC). J.W. is partly supported by IBS under the project IBS-R018-D3.

\clearpage

\begin{appendix}
    \section{Conventions and Useful Formulae}
    \label{Conventions}
    \subsection{Special-function conventions}\label{SpecialFunctions}
    We collect here the special-function conventions used throughout the paper.  Products and quotients of Gamma functions are abbreviated as
    \begin{align}
        \Gamma[a_1,\ldots,a_n]
        &\equiv
        \prod_{j=1}^{n}\Gamma(a_j),
        \\
        \Gamma\left[
        \begin{matrix}
            a_1,\ldots,a_n\\
            b_1,\ldots,b_m
        \end{matrix}
        \right]
        &\equiv
        \frac{\prod_{j=1}^{n}\Gamma(a_j)}
        {\prod_{\ell=1}^{m}\Gamma(b_\ell)}.
    \end{align}

    The Pochhammer symbol is
    \begin{align}
        (a)_z
        \equiv
        \frac{\Gamma(a+z)}{\Gamma(a)}.
    \end{align}
    The ordinary generalized hypergeometric function is defined, in the domain of convergence of its series, by
    \begin{align}
        {}_{p}\mathrm F_{q}\left[
        \begin{matrix}
            a_1,\ldots,a_p\\
            b_1,\ldots,b_q
        \end{matrix}
        \middle|z\right]
        \equiv
        \sum_{n=0}^{\infty}
        \frac{(a_1)_n\cdots(a_p)_n}
        {(b_1)_n\cdots(b_q)_n}
        \frac{z^n}{n!}.
        \label{eq:ordinary-hypergeometric}
    \end{align}
    The defining series assumes \(b_j\notin\mathbb Z_{\leq0}\).

    We also use the regularized and fully dressed variants
    \begin{align}
        {}_{p}\wh{\mathrm F}_{q}\left[
        \begin{matrix}a_{1},\ldots,a_{p}\\ b_{1},\ldots,b_{q}\end{matrix}
        \middle|z\right]
        &\equiv
        \frac{{}_{p}\mathrm F_{q}\left[
        \begin{matrix}a_{1},\ldots,a_{p}\\ b_{1},\ldots,b_{q}\end{matrix}
        \middle|z\right]}
        {\Gamma[b_{1},\ldots,b_{q}]},
        \\
        {}_{p}\mathcal F_{q}\left[
        \begin{matrix}a_{1},\ldots,a_{p}\\ b_{1},\ldots,b_{q}\end{matrix}
        \middle|z\right]
        &\equiv
        \Gamma\left[
        \begin{matrix}a_{1},\ldots,a_{p}\\ b_{1},\ldots,b_{q}\end{matrix}
        \right]
        {}_{p}\mathrm F_{q}\left[
        \begin{matrix}a_{1},\ldots,a_{p}\\ b_{1},\ldots,b_{q}\end{matrix}
        \middle|z\right].
        \label{eq:dressed-hypergeometric}
    \end{align}
    Outside their defining series domains, all three symbols denote analytic continuation in \(z\) and meromorphic continuation in their parameters. If a lower parameter lies in \(\mathbb Z_{\leq0}\), the regularized or dressed symbol denotes the analytic limit of the complete expression, rather than termwise substitution in the ordinary series.

    We also repeatedly use
    \begin{align}
        \operatorname*{Res}_{z=-n}\Gamma(z)
        =
        \frac{(-1)^n}{n!},
        \qquad n\in\mathbb N_0.
        \label{eq:gamma-pole-residue}
    \end{align}

    For fixed real \(x\), Stirling's formula on the principal branch gives the leading large-\(\lvert y\rvert\) behavior
    \begin{align}
        \Gamma\left(x+\I y\right)
        ={}&
        \sqrt{2\pi}\left|y\right|^{x-1/2}
        \E^{-\frac{\pi}{2}\left|y\right|}
        \nonumber\\
        &\times
        \exp\!\left\{
        \I\operatorname{sgn}(y)\left[
        \left|y\right|\bigl(\log\left|y\right|-1\bigr)
        +\frac{\pi}{2}\left(x-\frac12\right)
        \right]
        \right\}
        \nonumber\\
        &\times
        \left[1+\mathcal O\left(\lvert y\rvert^{-1}\right)\right],
        \qquad
        \left|y\right|\rightarrow\infty.
        \label{GammaLimit}
    \end{align}

    \subsection{Schwinger--Keldysh propagators and PMB representation}
    \label{MBr}
    A partial Mellin--Barnes (PMB) representation converts the Hankel functions in a dS propagator into powers of conformal time \cite{Qin:2022fbv,Qin:2022lva}.  This makes the vertex-time integrals algebraic in the Mellin variables and treats time derivatives systematically.

    Starting from the principal-series mode function \eqref{PriModeF}, with \(\wt\nu>0\), the two Wightman functions are
    \begin{align}
        \label{Dmp}
        D^{(\wt\nu)}_{-+}\left(k;\tau_1,\tau_2\right)
        ={}&
        \frac{\pi}{4}\E^{-\pi\wt\nu}
        (-\tau_1)^{3/2}(-\tau_2)^{3/2}
        \mathrm H_{\I\wt\nu}^{(1)}(-k\tau_1)
        \mathrm H_{-\I\wt\nu}^{(2)}(-k\tau_2),
        \\
        \label{Dpm}
        D^{(\wt\nu)}_{+-}\left(k;\tau_1,\tau_2\right)
        ={}&
        \frac{\pi}{4}\E^{-\pi\wt\nu}
        (-\tau_1)^{3/2}(-\tau_2)^{3/2}
        \mathrm H_{-\I\wt\nu}^{(2)}(-k\tau_1)
        \mathrm H_{\I\wt\nu}^{(1)}(-k\tau_2).
    \end{align}
    The two time-ordered propagators are
    \begin{align}
        \label{Dpmpm}
        D^{(\wt\nu)}_{\pm\pm}(k;\tau_1,\tau_2)
        ={}&
        D^{(\wt\nu)}_{\mp\pm}(k;\tau_1,\tau_2)
        \theta(\tau_1-\tau_2)
        +
        D^{(\wt\nu)}_{\pm\mp}(k;\tau_1,\tau_2)
        \theta(\tau_2-\tau_1).
    \end{align}
    On the principal-series line, \(D_{+-}=(D_{-+})^*\).  Complex conjugation is not, however, a holomorphic prescription away from real \(\wt\nu\). \eqref{Dmp} and \eqref{Dpm} therefore define the two Wightman branches independently before analytic continuation.  Throughout this paper, a physical complementary-series field is obtained by continuing the complete propagator according to
    \begin{align}
        \wt\nu\longrightarrow-\I\nu,
        \qquad 0<\nu<\frac32.
        \label{eq:propagator-mass-continuation}
    \end{align}
    In the mixed calculations below, this continuation is performed only after the relevant SK branches have been summed.

    For \(x>0\), the Mellin--Barnes representation of a Hankel function is
    \begin{align}
        \mathrm H_{\lambda}^{(j)}(x)
        =
        \int_{-\I\infty}^{+\I\infty}\frac{\D s}{2\pi\I}\,
        \frac{(x/2)^{-2s}}{\pi}
        \E^{(-1)^{j+1}(2s-\lambda-1)\frac{\I\pi}{2}}
        \Gamma\left[s-\frac{\lambda}{2},s+\frac{\lambda}{2}\right],
        \qquad j=1,2.
        \label{eq:hankel-mb}
    \end{align}
    The displayed integration path denotes a vertical Barnes contour whose constant real part satisfies \(\Re s>\lvert\Re\lambda\rvert/2\), so that all poles of the two Gamma functions lie to its left.  By \eqref{GammaLimit}, the representation is initially convergent for \(0<(-1)^{j+1}\arg x<\pi\).  Its values on the positive real axis are fixed by analytic continuation from this convergence domain.

    Define
    \begin{align}
        \mathcal G_{\wt\nu}(s)
        \equiv
        \Gamma\left[s-\frac{\I\wt\nu}{2},
                    s+\frac{\I\wt\nu}{2}\right].
        \label{eq:gamma-kernel}
    \end{align}
    Substituting \eqref{eq:hankel-mb} into \eqref{Dmp} and \eqref{Dpm}, the Wightman propagators become
    \begin{align}
        \label{PMBD}
        D^{(\wt\nu)}_{\pm\mp}(k;\tau_1,\tau_2)
        ={}&
        \frac{1}{4\pi}
        \int_{-\I\infty}^{+\I\infty}\frac{\D s_1}{2\pi\I}
        \int_{-\I\infty}^{+\I\infty}\frac{\D s_2}{2\pi\I}\,
        \E^{\mp\I\pi(s_1-s_2)}
        \left(\frac{k}{2}\right)^{-2s_{12}}
        \nonumber\\
        &\times
        (-\tau_1)^{-2s_1+3/2}
        (-\tau_2)^{-2s_2+3/2}
        \mathcal G_{\wt\nu}(s_1)\mathcal G_{\wt\nu}(s_2).
    \end{align}
    Combining \eqref{Dpmpm} with \eqref{PMBD} gives the time-ordered propagators
    \begin{align}
        \label{PMBDpp}
        D^{(\wt\nu)}_{\pm\pm}(k;\tau_1,\tau_2)
        ={}&
        \frac{1}{4\pi}
        \int_{-\I\infty}^{+\I\infty}\frac{\D s_1}{2\pi\I}
        \int_{-\I\infty}^{+\I\infty}\frac{\D s_2}{2\pi\I}\,
        \Big[
        \E^{\pm\I\pi(s_1-s_2)}\theta(\tau_1-\tau_2)
        +
        \E^{\mp\I\pi(s_1-s_2)}\theta(\tau_2-\tau_1)
        \Big]
        \left(\frac{k}{2}\right)^{-2s_{12}}
        \nonumber\\
        &\times
        (-\tau_1)^{-2s_1+3/2}
        (-\tau_2)^{-2s_2+3/2}
        \mathcal G_{\wt\nu}(s_1)\mathcal G_{\wt\nu}(s_2).
    \end{align}
    Here and below, the upper and lower signs are correlated throughout each expression.

    Before differentiating the Mellin integrand, consider the derivatives of the step functions in \eqref{Dpmpm}.  Their total contribution is
    \begin{align}
        &D^{(\wt\nu)}_{\mp\pm}(k;\tau_1,\tau_2)
        \partial_{\tau_2}\theta(\tau_1-\tau_2)
        +
        D^{(\wt\nu)}_{\pm\mp}(k;\tau_1,\tau_2)
        \partial_{\tau_2}\theta(\tau_2-\tau_1)
        \nonumber\\
        &\qquad=
        \delta(\tau_1-\tau_2)
        \left[
        D^{(\wt\nu)}_{\pm\mp}(k;\tau_1,\tau_2)
        -
        D^{(\wt\nu)}_{\mp\pm}(k;\tau_1,\tau_2)
        \right]
        =0,
    \end{align}
    because \eqref{Dmp} and \eqref{Dpm} coincide at \(\tau_1=\tau_2\).  Therefore, a single \(\tau_2\) derivative acts only on the power of \((-\tau_2)\):
    \begin{align}
        \partial_{\tau_2}
        (-\tau_2)^{-2s_2+3/2}
        =
        \left(2s_2-\frac32\right)
        (-\tau_2)^{-2s_2+1/2}.
    \end{align}
    Equivalently, the derivative multiplies the PMB integrand by \(2s_2-\frac32\) and lowers the power of \((-\tau_2)\) by one.  Applying this identity to \eqref{PMBD} and \eqref{PMBDpp} gives
    \begin{align}
        \label{PMBderivD}
        \partial_{\tau_2}
        D^{(\wt\nu)}_{\pm\mp}(k;\tau_1,\tau_2)
        ={}&
        \frac{1}{4\pi}
        \int_{-\I\infty}^{+\I\infty}\frac{\D s_1}{2\pi\I}
        \int_{-\I\infty}^{+\I\infty}\frac{\D s_2}{2\pi\I}\,
        \left(2s_2-\frac32\right)
        \E^{\mp\I\pi(s_1-s_2)}
        \left(\frac{k}{2}\right)^{-2s_{12}}
        \nonumber\\
        &\times
        (-\tau_1)^{-2s_1+3/2}
        (-\tau_2)^{-2s_2+1/2}
        \mathcal G_{\wt\nu}(s_1)\mathcal G_{\wt\nu}(s_2),
        \\
        \label{PMBderivDpp}
        \partial_{\tau_2}
        D^{(\wt\nu)}_{\pm\pm}(k;\tau_1,\tau_2)
        ={}&
        \frac{1}{4\pi}
        \int_{-\I\infty}^{+\I\infty}\frac{\D s_1}{2\pi\I}
        \int_{-\I\infty}^{+\I\infty}\frac{\D s_2}{2\pi\I}\,
        \left(2s_2-\frac32\right)
        \nonumber\\
        &\times
        \Big[
        \E^{\pm\I\pi(s_1-s_2)}\theta(\tau_1-\tau_2)
        +
        \E^{\mp\I\pi(s_1-s_2)}\theta(\tau_2-\tau_1)
        \Big]
        \left(\frac{k}{2}\right)^{-2s_{12}}
        \nonumber\\
        &\times
        (-\tau_1)^{-2s_1+3/2}
        (-\tau_2)^{-2s_2+1/2}
        \mathcal G_{\wt\nu}(s_1)\mathcal G_{\wt\nu}(s_2).
    \end{align}
    Higher derivatives of a time-ordered propagator can generate local contact terms at coincident times.

    \subsection{Barnes' first lemma}\label{BarnesFF}
    For complex parameters \(a_i\) and \(b_j\), Barnes' first lemma reads
    \begin{align}\label{BarnesLemma}
        &\int_{-\I\infty}^{+\I\infty}\frac{\D u}{2\pi\I}\,
        \Gamma(a_1+u)\Gamma(a_2+u)
        \Gamma(b_1-u)\Gamma(b_2-u)
        \nonumber\\
        &\hspace{1cm}=
        \Gamma\left[
        \begin{matrix}
        a_1+b_1,a_1+b_2,a_2+b_1,a_2+b_2\\
        a_1+a_2+b_1+b_2
        \end{matrix}
        \right].
    \end{align}
    \eqref{BarnesLemma} is understood by meromorphic continuation outside its initial convergence domain.

    \section{The Single-Exchange Tree Seed}\label{AppSingleExchange}
    \subsection{Exact principal-series result}
    This appendix collects the exact, SK-summed single-exchange seed \(\mathcal I^{p_1p_2}(r_1,r_2)\) defined in \eqref{eq_Idef}; its momentum ratios and relation to the physical trispectrum are given in \eqref{eq_TvsI}.  We first take the principal-series mass parameter \(\wt\nu\) defined in \eqref{PriModeF} and use the PMB propagators \eqref{PMBD} and \eqref{PMBDpp} \cite{Qin:2022fbv}.  The powers \(p_1\) and \(p_2\) encode the time dependence of the two vertices.  The time integral \eqref{eq_Idef} is first evaluated in its convergence domain, \(\operatorname{Re}p_{1,2}>-5/2\), and is defined elsewhere by meromorphic continuation of the complete result.  We use the Gamma-product, Pochhammer, and hypergeometric conventions of App.~\ref{SpecialFunctions}, in particular \eqref{eq:ordinary-hypergeometric} and \eqref{eq:dressed-hypergeometric}.

We first work with \(k_{12}>k_{34}\), or equivalently \(0<r_1<r_2<1\), where \(r_1=k_s/k_{12}\) and \(r_2=k_s/k_{34}\).  In this configuration, the exact result decomposes into a local signal, a nonlocal signal, and a background:
\begin{align}
\mathcal I^{p_1p_2}(r_1,r_2)= \mathcal I^{p_1p_2}_{\mathrm{LS}}(r_1,r_2) + \mathcal I^{p_1p_2}_{\mathrm{NS}}(r_1,r_2) + \mathcal I^{p_1p_2}_{\mathrm{BG}}(r_1,r_2).
\label{eq:single-exchange-seed-decomposition}
\end{align}
For integer \(p_1\) and \(p_2\), the local signal oscillates in \(\log(r_1/r_2)\).  After the conventional factor \(k_s^{5+p_{12}}\) is stripped off, it has an expansion in nonnegative integer powers of \(k_s^2\) at fixed external energies.  It is therefore local in the exchanged momentum:
    \begin{align}\label{LS1}
        \mathcal{I}^{p_1p_2}_{\mathrm{LS}}(r_{1},r_{2})
        =&~\FR{\cos\frac\pi2(p_1-p_2) + \sin \frac\pi2(p_{12}+2\I\wt\nu) }{2\pi}r_1^{5/2+p_1}r_2^{5/2+p_2}\Big(\FR{r_1}{r_2}\Big)^{\I\wt\nu}\n\\
       &\times \Gamma\Big[\FR52+p_1+\I\wt\nu,\FR52+p_2-\I\wt\nu,-\I\wt\nu,\I\wt\nu\Big]\n\\
        &\times {}_2\mathrm F_1\left[\begin{matrix}
        \frac54+\frac{p_1}2+\frac{\I\wt\nu}2,\frac74+\frac{p_1}2+\frac{\I\wt\nu}2\\1+\I\wt\nu
        \end{matrix}\middle| r_1^2\right]
         {}_2\mathrm F_1\left[\begin{matrix}
        \frac54+\frac{p_2}2-\frac{\I\wt\nu}2,\frac74+\frac{p_2}2-\frac{\I\wt\nu}2\\1-\I\wt\nu
        \end{matrix}\middle| r_2^2\right] + (\wt\nu\to-\wt\nu).
    \end{align}
By contrast, the nonlocal signal oscillates in \(\log(r_1r_2)\) and is nonanalytic in \(k_s\):
    \begin{align}\label{NS1}
        \mathcal{I}^{p_1p_2}_{\mathrm{NS}}(r_{1},r_{2})
        =&~\FR{\cos\frac\pi2(p_1-p_2) + \sin \frac\pi2(p_{12}+2\I\wt\nu) }{2\pi}r_1^{5/2+p_1}r_2^{5/2+p_2}\Big(\FR{r_1r_2}{4}\Big)^{\I\wt\nu}\n\\
        &\times \Gamma\Big[\FR52+p_1+\I\wt\nu,\FR52+p_2+\I\wt\nu,-\I\wt\nu,-\I\wt\nu\Big]\n\\
        &\times {}_2\mathrm F_1\left[\begin{matrix}
        \frac54+\frac{p_1}2+\frac{\I\wt\nu}2,\frac74+\frac{p_1}2+\frac{\I\wt\nu}2\\1+\I\wt\nu
        \end{matrix}\middle| r_1^2\right]
         {}_2\mathrm F_1\left[\begin{matrix}
        \frac54+\frac{p_2}2+\frac{\I\wt\nu}2,\frac74+\frac{p_2}2+\frac{\I\wt\nu}2\\1+\I\wt\nu
        \end{matrix}\middle| r_2^2\right] + (\wt\nu\to-\wt\nu).
    \end{align}
The background likewise contains no mass-dependent noninteger power of the exchanged momentum.  For the physical trispectrum \(p_1=p_2=0\), restoring the \(k_s^{-5}\) prefactor in \eqref{eq_TvsI} gives a background analytic in \(k_s^2\) at fixed external energies.  For \(r_1<r_2\), its seed admits the following series representation:
    \begin{align}
        \label{BG1}
        \mathcal{I}^{p_1p_2}_{\mathrm{BG}}(r_{1},r_{2})
        =&~ \FR{\cos(\frac\pi2p_{12})}{\I\wt\nu}r_1^{5+p_{12}}\sum_{n_1,n_2=0}^\infty \FR{(-1)^{n_{12}}}{n_1!n_2!}\Big(\FR{r_1}2\Big)^{2n_{12}}(-\I\wt\nu)_{-n_1}(\I\wt\nu)_{-n_2} \n\\
        & \times        {}_{2}\mathcal F_{1}\left[
\begin{matrix}
2n_{2}+p_2+\frac52-\I\wt\nu,2n_{12}+p_{12}+5 \\ 2n_{2}+p_2+\frac72-\I\wt\nu
\end{matrix}\middle| -\FR{r_1}{r_2}\right] + (\wt\nu\to-\wt\nu).
    \end{align}
For \(0<r_2<r_1<1\), \eqref{LS1} and \eqref{BG1} follow by \((r_1,p_1)\leftrightarrow(r_2,p_2)\), while \eqref{NS1} is symmetric under this interchange, as is also manifest from \eqref{eq_Idef}.  The limit \(\wt\nu\to0\) is taken only after the complete \(\mathcal I_{\mathrm{LS}}+\mathcal I_{\mathrm{NS}}\) result has been combined with all terms related by \(\wt\nu\to-\wt\nu\).

\subsection{Hierarchical squeezed limit}

For \(p_1=p_2=0\), the leading terms in the hierarchical squeezed limit \(r_1\ll r_2\ll1\) are
    \begin{align}
        \label{HSQLSSE}
        \lim_{r_1\ll r_2\ll 1}\mathcal{I}^{00}_{\mathrm{LS}}(r_{1},r_{2})
        =&~
       \FR{1+\I \sinh \pi\wt\nu}{2\pi}(r_1r_2)^{5/2}\Big(\FR{r_1}{r_2}\Big)^{\I\wt\nu}
        \Gamma\Big[\FR52+\I\wt\nu,\FR52-\I\wt\nu,-\I\wt\nu,\I\wt\nu\Big] + (\wt\nu\to-\wt\nu),\\
        \label{HSQNSSE}
        \lim_{r_1\ll r_2\ll 1}\mathcal{I}^{00}_{\mathrm{NS}}(r_{1},r_{2})
        =&~
        \FR{1+\I \sinh \pi\wt\nu}{2\pi}(r_1r_2)^{5/2}\Big(\FR{r_1r_2}{4}\Big)^{\I\wt\nu}
        \Gamma\Big[\FR52+\I\wt\nu,-\I\wt\nu\Big]^2 + (\wt\nu\to-\wt\nu),\\
        \label{HSQBGSE}
        \lim_{r_1\ll r_2\ll 1}\mathcal{I}^{00}_{\mathrm{BG}}(r_{1},r_{2})
        =&~\FR{192}{25+4\wt\nu^2}r_1^5.
    \end{align}
The sum of \eqref{HSQLSSE} and \eqref{HSQNSSE} reproduces \eqref{eq_PriIs}, whereas the analytic background \eqref{HSQBGSE} starts at order \(r_1^5\) and is subleading.  A physical complementary-series exchange is obtained by applying the continuation \eqref{eq:propagator-mass-continuation} to the complete result.  The branch proportional to \((r_1r_2)^{-\nu}\) then dominates the nonlocal signal in the hierarchical squeezed limit and gives \eqref{eq_ComIs}.

\subsection{Folded limit}

The second configuration needed in the main text is the folded limit \(r_2\to1^-\) of the bispectrum seed in \eqref{eq:single-exchange-bispectrum-seed}, for which \(r=r_1\) and \(0<r<1\).  At fixed SK indices \((\mathsf a,\mathsf b)\), separate terms in an intermediate representation can be singular in this limit; for the bispectrum seed \(p_2=-2\), the singularities are logarithmic.  The spurious divergences cancel within each individual SK branch as a consequence of the BD initial condition, so no cancellation between different branches is required. After taking these finite branchwise limits, summing the four SK branches gives the following expression obtained with the improved bootstrap method of \cite{Qin:2023ejc}:
\begin{align}\label{eq:folded-seed-general}
\mathcal I^{p_1p_2}(r,1) =&~ \frac{\cos\frac\pi2(p_1-p_2) + \sin \frac\pi2(p_{12}+2\I\wt\nu) }{2^{5/2+p_2+\I\wt\nu}\pi^{1/2}}\Big(\frac{r}{1+r}\Big)^{5/2+p_1+\I\wt\nu}
\Gamma\Big[ \frac{5}{2}+p_1+\I\wt\nu,-\I\wt\nu \Big] \n\\
&\times
\Gamma\left[
\begin{matrix}
\frac52+p_2-\I\wt\nu,\frac52+p_2+\I\wt\nu\\
3+p_2
\end{matrix}
\right]\times {}_2\mathrm F_1\left[
\begin{matrix}
\frac52+p_1+\I\wt\nu,\frac12+\I\wt\nu\\
1+2\I\wt\nu
\end{matrix}\middle| \frac{2r}{1+r}
\right] + (\wt\nu\to-\wt\nu) \n\\
&+ \frac{2\cos(\frac\pi2p_{12})}{(\frac52+p_2)^2+\wt\nu^2} \Big(\frac{r}{1+r}\Big)^{5+p_{12}}\Gamma(5+p_{12})\times {}_3\mathrm F_2\left[
\begin{matrix}
1,3+p_2,5+p_{12}\\ \frac72+p_2-\I\wt\nu,\frac72+p_2+\I\wt\nu
\end{matrix}\middle| \frac{2r}{1+r}
\right].
\end{align}
For the bispectrum seed \(p_1=0\), \(p_2=-2\), \eqref{eq:folded-seed-general} reduces to
\begin{align}\label{eq_foldedI}
\mathcal I^{0,-2}(r,1)
=& -\bigg\{
\frac{1+ \I\sinh\pi\wt\nu}
{2^{1/2+\I\wt\nu}\pi^{1/2}}
\Big(\frac{r}{1+r}\Big)^{5/2+\I\wt\nu}
\nonumber\\
&\times
\Gamma\Big[
 -\I\wt\nu,\frac12-\I\wt\nu,1+2\I\wt\nu
\Big]
{}_2\mathcal F_1\left[
\begin{matrix}
\frac12+\I\wt\nu,\frac52+\I\wt\nu\\
1+2\I\wt\nu
\end{matrix}\middle| \frac{2r}{1+r}
\right] \n\\
& + (\wt\nu\to-\wt\nu) \bigg\}
\nonumber\\
& - \frac{4}{\wt\nu^2+1/4} \Big(\frac{r}{1+r}\Big)^{3} {}_3\mathrm F_2\left[
\begin{matrix}
1,1,3\\ \frac32-\I\wt\nu,\frac32+\I\wt\nu
\end{matrix}\middle| \frac{2r}{1+r}
\right].
\end{align}
The complementary-series result follows by applying \eqref{eq:propagator-mass-continuation} to the complete folded expression.

\section{Tree Diagram with Two General Principal-Series Scalars}
\label{CalTree}

Following the mass convention in \eqref{PriModeF}, this appendix computes the tree diagram in Fig.~\ref{FigTree2pt} with two independent principal-series parameters \(\widetilde\nu_i=\sqrt{m_i^2-9/4}\), \(i=1,2\), initially taken in \(\mathbb R_{+}\).  By convention, the derivative at the mixing vertex acts on the line carrying \(\widetilde\nu_1\).  The physical continuation is performed only after the full general-mass result has been obtained.

In the orientation \(1|2\), the line with parameter \(\widetilde\nu_1\) joins the left endpoint to the mixing vertex, the line with parameter \(\widetilde\nu_2\) joins the mixing vertex to the right endpoint, and the derivative acts on the \(\widetilde\nu_1\) line.  We define
\begin{align}
\mathcal K_{1|2}(r_1,r_2)
={}&
(k_{12}k_{34})^{5/2}
\sum_{\mathsf a,\mathsf b,\mathsf c=\pm}
(-\I\mathsf a)(-\I\mathsf b)(-\I\mathsf c)
\int_{-\infty}^{0}
\D\tau_L\D\tau_M\D\tau_R\,
(-\tau_M)^{-3}
\nonumber\\
&\times
\E^{\I\mathsf a k_{12}\tau_L+\I\mathsf c k_{34}\tau_R}
\left[
\partial_{\tau_M}
D_{\mathsf a\mathsf b}^{(\widetilde\nu_1)}
(k_s;\tau_L,\tau_M)
\right]
D_{\mathsf b\mathsf c}^{(\widetilde\nu_2)}
(k_s;\tau_M,\tau_R).
\label{eq:tree-K-12}
\end{align}
In the opposite orientation \(2|1\), the two masses are exchanged between the endpoint lines while the derivative continues to act on the \(\widetilde\nu_1\) line:
\begin{align}
\mathcal K_{2|1}(r_1,r_2)
={}&
(k_{12}k_{34})^{5/2}
\sum_{\mathsf a,\mathsf b,\mathsf c=\pm}
(-\I\mathsf a)(-\I\mathsf b)(-\I\mathsf c)
\int_{-\infty}^{0}
\D\tau_L\D\tau_M\D\tau_R\,
(-\tau_M)^{-3}
\nonumber\\
&\times
\E^{\I\mathsf a k_{12}\tau_L+\I\mathsf c k_{34}\tau_R}
D_{\mathsf a\mathsf b}^{(\widetilde\nu_2)}
(k_s;\tau_L,\tau_M)
\left[
\partial_{\tau_M}
D_{\mathsf b\mathsf c}^{(\widetilde\nu_1)}
(k_s;\tau_M,\tau_R)
\right].
\label{eq:tree-K-21}
\end{align}
Each signed interaction in \eqref{Lagrangian2pt} contributes \(-\I\mathsf a\), \(-\I\mathsf b\), or \(-\I\mathsf c\).  Thus the factor in both seed definitions is \((-\I\mathsf a)(-\I\mathsf b)(-\I\mathsf c) =\I\mathsf a\mathsf b\mathsf c\).

The physical sum and exact exchange relation are given in \eqref{TreeSeed}.  Both seeds are evaluated in the configuration \(0<r_1<r_2<1\), before implementing the continuation in \eqref{eq:tree-physical-continuation}.

\subsection{Eight SK assignments and a hierarchy-adapted time basis}

We keep all eight SK assignments explicit.  This is important for the later continuation in \(\widetilde\nu_1\): complex conjugation is not a holomorphic operation on a complex mass parameter.  In the configuration \(0<r_1<r_2<1\), we decompose the product of the two propagators in each orientation in the hierarchy-adapted basis
\begin{align}
 1,\qquad
 \theta(\tau_M-\tau_L),\qquad
 \theta(\tau_M-\tau_R),\qquad
 \theta(\tau_R-\tau_M)\theta(\tau_M-\tau_L).
 \label{eq:tree-time-basis}
\end{align}
This organization of the time orderings into factorized and nested structures is consistent with the family-tree decomposition of nested time integrals \cite{Xianyu:2023ytd,Fan:2024iek}. The basis is positional and does not depend on which mass occupies the left or right propagator; it is therefore used for both \(\mathcal K_{1|2}\) and \(\mathcal K_{2|1}\). For the \(1|2\) orientation, define the two Wightman differences
\begin{align}
 \Delta^{(1)}_{\mathsf b}
 &\equiv
 \left[
 D^{(\widetilde\nu_1)}_{\mathsf b,-\mathsf b}
 -D^{(\widetilde\nu_1)}_{-\mathsf b,\mathsf b}
 \right](k_s;\tau_L,\tau_M),
 \nonumber\\
 \Delta^{(2)}_{\mathsf b}
 &\equiv
 \left[
 D^{(\widetilde\nu_2)}_{-\mathsf b,\mathsf b}
 -D^{(\widetilde\nu_2)}_{\mathsf b,-\mathsf b}
 \right](k_s;\tau_M,\tau_R).
 \label{eq:tree-wightman-differences}
\end{align}
The doubly nested weight in the usual propagator decomposition is rewritten in this basis using the exact identity
\begin{align}
 \theta(\tau_M-\tau_L)\theta(\tau_M-\tau_R)
 ={}&
 \theta(\tau_M-\tau_L)
 -
 \theta(\tau_R-\tau_M)\theta(\tau_M-\tau_L).
 \label{eq:tree-theta-change-of-basis}
\end{align}
After the change of basis \eqref{eq:tree-theta-change-of-basis}, the product of propagators becomes
\begin{align}
 &D^{(\widetilde\nu_1)}_{\mathsf a\mathsf b}
 (k_s;\tau_L,\tau_M)
 D^{(\widetilde\nu_2)}_{\mathsf b\mathsf c}
 (k_s;\tau_M,\tau_R)
 \nonumber\\
 ={}&
 D^{(\widetilde\nu_1)}_{-\mathsf b,\mathsf b}
 (k_s;\tau_L,\tau_M)
 D^{(\widetilde\nu_2)}_{\mathsf b,-\mathsf b}
 (k_s;\tau_M,\tau_R)
 \nonumber\\
 &+
 \delta_{\mathsf a,\mathsf b}\,
 \Delta^{(1)}_{\mathsf b}
 D^{(\widetilde\nu_2)}_{-\mathsf c,\mathsf c}
 (k_s;\tau_M,\tau_R)
 \theta(\tau_M-\tau_L)
 \nonumber\\
 &+
 \delta_{\mathsf b,\mathsf c}\,
 D^{(\widetilde\nu_1)}_{-\mathsf b,\mathsf b}
 (k_s;\tau_L,\tau_M)
 \Delta^{(2)}_{\mathsf b}
 \theta(\tau_M-\tau_R)
 \nonumber\\
 &-
 \delta_{\mathsf a,\mathsf b}
 \delta_{\mathsf b,\mathsf c}\,
 \Delta^{(1)}_{\mathsf b}
 \Delta^{(2)}_{\mathsf b}
 \theta(\tau_R-\tau_M)\theta(\tau_M-\tau_L).
 \label{eq:tree-hierarchy-adapted-product}
\end{align}
\eqref{eq:tree-hierarchy-adapted-product} is an identity before any time or Mellin integration.  Its \(\theta(\tau_M-\tau_L)\) coefficient combines the original left-nested term with the reassigned part of the doubly nested term.

When this identity is inserted into \eqref{eq:tree-K-12}, \(\partial_{\tau_M}\) acts only on the mass-\(\widetilde\nu_1\) object in each line, namely \(D^{(\widetilde\nu_1)}\) or \(\Delta^{(1)}\); it does not act on the mass-\(\widetilde\nu_2\) object or the rewritten time weights.

The four lines on the right-hand side of \eqref{eq:tree-hierarchy-adapted-product} follow the order in \eqref{eq:tree-time-basis}.  We label them by \([\mathrm{FF}]\), \([\mathrm{NF}]\), \([\mathrm{FN}]\), and \([\mathrm{NN}]\), respectively.  Tab.~\ref{tab:eight-sk-tree} lists the SK assignments that support each contribution.
\begin{table}[t]
\centering
\begin{tabular}{c|c}
\toprule
\((\mathsf a,\mathsf b,\mathsf c)\)
& hierarchy-adapted contributions
\\
\midrule
\((-,+,-)\) & \([\mathrm{FF}]\)\\
\((+,-,+)\) & \([\mathrm{FF}]\)\\
\((+,+,-)\) & \([\mathrm{FF}],[\mathrm{NF}]\)\\
\((-,-,+)\) & \([\mathrm{FF}],[\mathrm{NF}]\)\\
\((-,+,+)\) & \([\mathrm{FF}],[\mathrm{FN}]\)\\
\((+,-,-)\) & \([\mathrm{FF}],[\mathrm{FN}]\)\\
\((+,+,+)\) &
\([\mathrm{FF}],[\mathrm{NF}],[\mathrm{FN}],[\mathrm{NN}]\)\\
\((-,-,-)\) &
\([\mathrm{FF}],[\mathrm{NF}],[\mathrm{FN}],[\mathrm{NN}]\)\\
\bottomrule
\end{tabular}
\caption{The eight SK assignments and the hierarchy-adapted contributions
present in each branch.}
\label{tab:eight-sk-tree}
\end{table}

After all eight rows are summed, each orientation contains exactly four contributions in the same order:
\begin{align}
 \mathcal K_{1|2}
 =
 \mathcal K_{1|2}^{[\mathrm{FF}]}
 +
 \mathcal K_{1|2}^{[\mathrm{NF}]}
 +
 \mathcal K_{1|2}^{[\mathrm{FN}]}
 +
 \mathcal K_{1|2}^{[\mathrm{NN}]}.
 \label{eq:four-factorization-structures}
 \\
 \mathcal K_{2|1}
 =
 \mathcal K_{2|1}^{[\mathrm{FF}]}
 +
 \mathcal K_{2|1}^{[\mathrm{NF}]}
 +
 \mathcal K_{2|1}^{[\mathrm{FN}]}
 +
 \mathcal K_{2|1}^{[\mathrm{NN}]}.
 \label{eq:four-factorization-structures-21}
\end{align}
Thus \([\mathrm{FF}]\) is fully factorized, \([\mathrm{NF}]\) and \([\mathrm{FN}]\) are the two partially nested contributions, and \([\mathrm{NN}]\) is the ordered fully nested chain \(\tau_L<\tau_M<\tau_R\).  The labels refer to the four terms in the exact reorganization \eqref{eq:tree-hierarchy-adapted-product}; they do not mean that the two internal propagators were factorized independently after the change of basis. As shown in App.~\ref{MBr}, differentiating a time-ordered propagator produces no additional contribution: the derivative of the step function is proportional to the difference of two equal-time Wightman functions and therefore vanishes.  Hence the derivative acts only on the explicit power of \((-\tau_M)\), as in \eqref{PMBderivDpp}.

\subsection{Mellin--Barnes evaluation of the oriented seeds}

We use the special-function conventions of App.~\ref{SpecialFunctions} and the Gamma kernel \(\mathcal G_{\widetilde\nu_i}(s)\) defined in \eqref{eq:gamma-kernel}.  For \(z,z'>0\), rescaling the Wightman representation \eqref{PMBD} gives
\begin{align}
 k_s^3D^{(\widetilde\nu_i)}_{\mathsf a,-\mathsf a}
 \left(k_s;-\frac z{k_s},-\frac{z'}{k_s}\right)
 ={}&
 \frac1{4\pi}
 \int_{-\I\infty}^{+\I\infty}\frac{\D s}{2\pi\I}
 \int_{-\I\infty}^{+\I\infty}\frac{\D t}{2\pi\I}\,
 2^{2(s+t)}
 z^{3/2-2s}(z')^{3/2-2t}
 \nonumber\\
 &\times
 \mathcal G_{\widetilde\nu_i}(s)
 \mathcal G_{\widetilde\nu_i}(t)
 \E^{-\I\pi\mathsf a(s-t)}.
 \label{eq:tree-pmb-wightman}
\end{align}
We label the four Mellin variables for \(1|2\), ordered from the left endpoint to the right endpoint, by \((s_1,s_2;s_3,s_4)\).  Thus \((s_1,s_2)\) belongs to the propagator with mass \(\widetilde\nu_1\), while \((s_3,s_4)\) belongs to the propagator with mass \(\widetilde\nu_2\).  The dimensionless vertex times are \(z_L=-k_s\tau_L\), \(z_M=-k_s\tau_M\), and \(z_R=-k_s\tau_R\).

We define the Mellin measure by
\begin{align}
 [\D s]
 \equiv
 \prod_{i=1}^{4}\frac{\D s_i}{2\pi\I}.
 \label{eq:tree-mellin-measure}
\end{align}
We use the contracted-subscript convention of the main text.  With the vertex weights included, the three time integrals have the power-law dependence \(z_L^{A_L-1}z_M^{A_M-1}z_R^{A_R-1}\), where
\begin{align}
 A_L&=\frac52-2s_1,
 &
 A_M&=\delta-2s_{23},
 &
 A_R&=\frac52-2s_4.
 \label{eq:tree-time-exponents}
\end{align}
Here \(\delta\) is a temporary analytic regulator defined by the dimensionless insertion
\begin{align}
 (-\tau_M)^{-3}
 \longrightarrow
 (-\tau_M)^{-3}(-k_s\tau_M)^{\delta}.
 \label{eq:tree-middle-time-regulator}
\end{align}
The three nested sectors are first evaluated for sufficiently large \(\operatorname{Re}\delta\), where the propagator and time-kernel poles can be separated, while the fully factorized middle-time integral is defined distributionally by the Mellin delta with the same regulator.  The regulated expressions are then continued to \(\delta=0\), with the time-kernel divisors kept on the side opposing the propagator poles.  Every unregulated formula below means this prescribed limit.  This middle-time regulator defines only the meromorphic continuation and is unrelated to the convergent contour selected by an endpoint SK phase. Each oscillatory integral over \(z_L,z_M,z_R>0\) is evaluated on the convergent contour selected by its SK phase.  This uniquely fixes all complex powers.  For \(r_i>0\), \(i=1,2\), and \(\mathsf a=\pm1\), we use
\begin{align}
 \left(\frac{\I\mathsf a}{r_i}\right)^{-A}
 =
 (-\I\mathsf a r_i)^A
 =
 r_i^A\exp\left(-\frac{\I\pi\mathsf a A}{2}\right).
\end{align}
By \eqref{PMBderivDpp}, the derivative on the second time argument of the left propagator supplies the common factor \(2s_2-3/2\).  We define the four time kernels directly by their integration domains:
\begin{align}
 \mathfrak T_{\mathsf a\mathsf b\mathsf c}^{[\mathrm{FF}]}
 (A_L,A_M,A_R)
 ={}&
 \int_0^\infty\D z_L\,z_L^{A_L-1}
 \E^{-\I\mathsf a z_L/r_1}
 \int_0^\infty\D z_M\,z_M^{A_M-1}
 \int_0^\infty\D z_R\,z_R^{A_R-1}
 \E^{-\I\mathsf c z_R/r_2},
 \label{eq:time-domain-ff}
 \\
 \mathfrak T_{\mathsf b\mathsf b\mathsf c}^{[\mathrm{NF}]}
 (A_L,A_M,A_R)
 ={}&
 \int_0^\infty\D z_L\,z_L^{A_L-1}
 \E^{-\I\mathsf b z_L/r_1}
 \int_0^{z_L}\D z_M\,z_M^{A_M-1}
 \int_0^\infty\D z_R\,z_R^{A_R-1}
 \E^{-\I\mathsf c z_R/r_2},
 \label{eq:time-domain-nf}
 \\
 \mathfrak T_{\mathsf a\mathsf b\mathsf b}^{[\mathrm{FN}]}
 (A_L,A_M,A_R)
 ={}&
 \int_0^\infty\D z_L\,z_L^{A_L-1}
 \E^{-\I\mathsf a z_L/r_1}
 \int_0^\infty\D z_R\,z_R^{A_R-1}
 \E^{-\I\mathsf b z_R/r_2}
 \int_0^{z_R}\D z_M\,z_M^{A_M-1},
 \label{eq:time-domain-fn}
 \\
 \mathfrak T_{\mathsf b\mathsf b\mathsf b}^{[\mathrm{NN}]}
 (A_L,A_M,A_R)
 ={}&
 \int_0^\infty\D z_L\,z_L^{A_L-1}
 \E^{-\I\mathsf b z_L/r_1}
 \int_0^{z_L}\D z_M\,z_M^{A_M-1}
 \int_0^{z_M}\D z_R\,z_R^{A_R-1}
 \E^{-\I\mathsf b z_R/r_2}.
 \label{eq:time-domain-nn}
\end{align}
The last line follows from \(\theta(\tau_R-\tau_M)\theta(\tau_M-\tau_L)\), which becomes \(0<z_R<z_M<z_L\).

Substituting \eqref{eq:tree-pmb-wightman} into \eqref{eq:tree-K-12} and using the four time domains \eqref{eq:time-domain-ff}-- \eqref{eq:time-domain-nn} gives four formulas, the first one is
\begin{align}
 \mathcal K_{1|2}^{[\mathrm{FF}]}
 ={}&
 \frac{\I(r_1r_2)^{-5/2}}{16\pi^2}
 \sum_{\mathsf a,\mathsf b,\mathsf c=\pm}
 \mathsf a\mathsf b\mathsf c
 \int[\D s]\,
 2^{2s_{1234}}
 \left(2s_2-\frac32\right)
 \nonumber\\
 &\times
 \prod_{i=1}^{2}\mathcal G_{\widetilde\nu_1}(s_i)
 \prod_{j=3}^{4}\mathcal G_{\widetilde\nu_2}(s_j)
 \nonumber\\
 &\times
 \E^{+\I\pi\mathsf b(s_1-s_2)}
 \E^{-\I\pi\mathsf b(s_3-s_4)}
 \mathfrak T_{\mathsf a\mathsf b\mathsf c}^{[\mathrm{FF}]}
 (A_L,A_M,A_R).
 \label{eq:tree-pmb-ff}
\end{align}
The contribution nested on the left has \(\mathsf a=\mathsf b\).  Because of the reorganization in \eqref{eq:tree-hierarchy-adapted-product}, its right Wightman function is \(D^{(\widetilde\nu_2)}_{-\mathsf c,\mathsf c}\), and hence
\begin{align}
 \mathcal K_{1|2}^{[\mathrm{NF}]}
 ={}&
 \frac{(r_1r_2)^{-5/2}}{8\pi^2}
 \sum_{\mathsf b,\mathsf c=\pm}
 \mathsf b\mathsf c
 \int[\D s]\,
 2^{2s_{1234}}
 \left(2s_2-\frac32\right)
 \nonumber\\
 &\times
 \prod_{i=1}^{2}\mathcal G_{\widetilde\nu_1}(s_i)
 \prod_{j=3}^{4}\mathcal G_{\widetilde\nu_2}(s_j)
 \sin[\pi(s_1-s_2)]
 \E^{+\I\pi\mathsf c(s_3-s_4)}
 \mathfrak T_{\mathsf b\mathsf b\mathsf c}^{[\mathrm{NF}]}
 (A_L,A_M,A_R).
 \label{eq:tree-pmb-nf}
\end{align}
Similarly, nesting on the right imposes \(\mathsf c=\mathsf b\) and gives
\begin{align}
 \mathcal K_{1|2}^{[\mathrm{FN}]}
 ={}&
 -\frac{(r_1r_2)^{-5/2}}{8\pi^2}
 \sum_{\mathsf a,\mathsf b=\pm}
 \mathsf a\mathsf b
 \int[\D s]\,
 2^{2s_{1234}}
 \left(2s_2-\frac32\right)
 \nonumber\\
 &\times
 \prod_{i=1}^{2}\mathcal G_{\widetilde\nu_1}(s_i)
 \prod_{j=3}^{4}\mathcal G_{\widetilde\nu_2}(s_j)
 \E^{+\I\pi\mathsf b(s_1-s_2)}
 \sin[\pi(s_3-s_4)]
 \mathfrak T_{\mathsf a\mathsf b\mathsf b}^{[\mathrm{FN}]}
 (A_L,A_M,A_R).
 \label{eq:tree-pmb-fn}
\end{align}
Finally, the ordered fully nested contribution has \(\mathsf a=\mathsf b=\mathsf c\).  The minus sign in the last line of \eqref{eq:tree-hierarchy-adapted-product}, combined with \((-\I\mathsf a)(-\I\mathsf b)(-\I\mathsf c) =\I\mathsf a\mathsf b\mathsf c\) in \eqref{eq:tree-K-12}, gives
\begin{align}
 \mathcal K_{1|2}^{[\mathrm{NN}]}
 ={}&
 -\frac{\I(r_1r_2)^{-5/2}}{4\pi^2}
 \sum_{\mathsf b=\pm}
 \mathsf b
 \int[\D s]\,
 2^{2s_{1234}}
 \left(2s_2-\frac32\right)
 \nonumber\\
 &\times
 \prod_{i=1}^{2}\mathcal G_{\widetilde\nu_1}(s_i)
 \prod_{j=3}^{4}\mathcal G_{\widetilde\nu_2}(s_j)
 \sin[\pi(s_1-s_2)]
 \sin[\pi(s_3-s_4)]
 \mathfrak T_{\mathsf b\mathsf b\mathsf b}^{[\mathrm{NN}]}
 (A_L,A_M,A_R).
 \label{eq:tree-pmb-nn}
\end{align}
The sine factors in \eqref{eq:tree-pmb-nf}--\eqref{eq:tree-pmb-nn} come from the Wightman differences in \eqref{eq:tree-wightman-differences}.  The time restrictions are the four domains in \eqref{eq:time-domain-ff}--\eqref{eq:time-domain-nn}.

\paragraph{Evaluation of the time kernels}

Unless stated otherwise, all ordinary time integrals below are evaluated in a common domain of absolute convergence, with their physical values obtained by meromorphic continuation and the limit \(\delta\to0\).  The zero-energy middle integral in the \([\mathrm{FF}]\) sector is instead understood distributionally as a Mellin delta.  Evaluating the time integrals defined in \eqref{eq:time-domain-ff}--\eqref{eq:time-domain-nn}, the fully factorized kernel gives
\begin{align}
 \mathfrak T_{\mathsf a\mathsf b\mathsf c}^{[\mathrm{FF}]}
 (A_L,A_M,A_R)
 ={}&
 \Gamma[A_L,A_R]\,
 (-\I\mathsf a r_1)^{A_L}
 (-\I\mathsf c r_2)^{A_R}
 2\pi\I\delta_{\mathrm M}(A_M),
 \label{eq:time-ff}
\end{align}
where the Mellin delta is normalized so that
\begin{align}
 \int\frac{\D s_3}{2\pi\I}\,
 2\pi\I\delta_{\mathrm M}
 \left(\delta-2s_{23}\right)f(s_3)
 =
 \frac12 f\left(\frac\delta2-s_2\right).
 \label{eq:mellin-delta-normalization}
\end{align}
The two partially nested kernels give
\begin{align}
 \mathfrak T_{\mathsf b\mathsf b\mathsf c}^{[\mathrm{NF}]}
 (A_L,A_M,A_R)
 ={}&
 \frac{\Gamma[A_L+A_M,A_R]}{A_M}
 (-\I\mathsf b r_1)^{A_L+A_M}
 (-\I\mathsf c r_2)^{A_R},
 \label{eq:time-nf}
 \\
 \mathfrak T_{\mathsf a\mathsf b\mathsf b}^{[\mathrm{FN}]}
 (A_L,A_M,A_R)
 ={}&
 \frac{\Gamma[A_L,A_R+A_M]}{A_M}
 (-\I\mathsf a r_1)^{A_L}
 (-\I\mathsf b r_2)^{A_R+A_M}.
 \label{eq:time-fn}
\end{align}
For the ordered fully nested contribution, let \(A_T=A_L+A_M+A_R=5+\delta-2s_{1234}\).  Expanding \eqref{eq:time-domain-nn} in \(r_1/r_2\) for \(r_1<r_2\) gives
\begin{align}
 \mathfrak T_{\mathsf b\mathsf b\mathsf b}^{[\mathrm{NN}]}
 (A_L,A_M,A_R)
 ={}&
 (-\I\mathsf b r_1)^{A_T}
 \sum_{m=0}^{\infty}
 \frac{\Gamma(A_T+m)}{m!(A_R+m)(A_R+A_M+m)}
 \left(-\frac{r_1}{r_2}\right)^m.
 \label{eq:time-nn-series}
\end{align}
Using \eqref{eq:ordinary-hypergeometric} and \eqref{eq:dressed-hypergeometric}, the series \eqref{eq:time-nn-series} sums to
\begin{align}
 \mathfrak T_{\mathsf b\mathsf b\mathsf b}^{[\mathrm{NN}]}
 (A_L,A_M,A_R)
 ={}&
 \frac{\Gamma(A_T)}{A_R(A_R+A_M)}
 (-\I\mathsf b r_1)^{A_T}
 {}_3\mathrm F_2\left[
 \begin{matrix}
 A_T,A_R,A_R+A_M\\
 A_R+1,A_R+A_M+1
 \end{matrix}
 \middle|
 -\frac{r_1}{r_2}
 \right]
 \nonumber\\
 ={}&
 (-\I\mathsf b r_1)^{A_T}
 {}_3\mathcal F_2\left[
 \begin{matrix}
 A_T,A_R,A_R+A_M\\
 A_R+1,A_R+A_M+1
 \end{matrix}
 \middle|
 -\frac{r_1}{r_2}
 \right].
 \label{eq:time-nn}
\end{align}
\eqref{eq:tree-pmb-ff}--\eqref{eq:tree-pmb-nn} and \eqref{eq:time-ff}--\eqref{eq:time-nn} constitute a complete analytic representation of all eight SK assignments of \(\mathcal K_{1|2}\) in the basis \eqref{eq:tree-time-basis}.

\subsection{Residue evaluation for the hierarchical configuration}

For the hierarchical configuration \(0<r_1<r_2<1\), the Mellin--Barnes expansion appropriate to this hierarchy is generated by the left residue cone. Mellin-space time integrations, including arbitrarily nested ones, generate only right poles \cite{Qin:2023bjk,Xianyu:2023ytd}; the explicit kernels \eqref{eq:time-ff}--\eqref{eq:time-nn} provide a direct check in the present case.  Hence no ordinary time-integral pole belongs to this cone.  The Mellin delta in the \([\mathrm{FF}]\) sector is instead imposed according to \eqref{eq:mellin-delta-normalization} and treated separately in the fully factorized contribution below.  Thus the only ordinary left poles are those of the Gamma factors \(\mathcal G_{\widetilde\nu_i}\) defined in \eqref{eq:gamma-kernel}.

For \(\mathcal K_{1|2}\), these poles are
\begin{align}
 s_i
 &=
 -n_i-\frac{\I\dd_i\widetilde\nu_1}{2},
 &&i=1,2,
 \nonumber\\
 s_i
 &=
 -n_i-\frac{\I\dd_i\widetilde\nu_2}{2},
 &&i=3,4.
 \label{eq:tree-pole-convention}
\end{align}
Here \(\dd_i=\pm1\) and \(n_i\in\mathbb N_0\).  The mode-function residues at \eqref{eq:tree-pole-convention} follow from \eqref{eq:gamma-pole-residue}.  At simultaneous residues of both propagators, the nonvanishing sine factors require
\begin{align}
 [\mathrm{NF}]:\quad&
 \dd_1=-\dd_2,
 \nonumber\\
 [\mathrm{FN}]:\quad&
 \dd_3=-\dd_4,
 \nonumber\\
 [\mathrm{NN}]:\quad&
 \dd_1=-\dd_2,
 \qquad
 \dd_3=-\dd_4.
 \label{eq:nested-branch-selection}
\end{align}
Equal branch signs make the corresponding sine vanish. In the \([\mathrm{FF}]\) sector, the endpoint branches \(\dd_1,\dd_4=\pm1\) are independent, while the two central Mellin variables are tied by \(\delta_{\mathrm M}(A_M)\).

\paragraph{The partially and fully nested contributions}

Taking the propagator residues in \eqref{eq:tree-pmb-nf} and \eqref{eq:tree-pmb-fn}, the complete SK sums give
\begin{align}
 \mathcal K_{1|2}^{[\mathrm{NF}]}(r_1,r_2)
 ={}&
 \frac{\I\sinh(\pi\widetilde\nu_1)}{2\pi^2}
 \sum_{\dd_1,\dd_3,\dd_4=\pm1}
 \sum_{n_1,n_2,n_3,n_4=0}^{\infty}
 \dd_1
 \frac{(-1)^{n_{1234}}}{n_1!n_2!n_3!n_4!}
 \nonumber\\
 &\times
 \Gamma\left[
 -n_1-\I\dd_1\widetilde\nu_1,
 -n_2+\I\dd_1\widetilde\nu_1,
 -n_3-\I\dd_3\widetilde\nu_2,
 -n_4-\I\dd_4\widetilde\nu_2
 \right]
 \nonumber\\
 &\times
 \frac{-2n_2-\frac32+\I\dd_1\widetilde\nu_1}
 {2n_{23}+\I(\dd_3\widetilde\nu_2-
                  \dd_1\widetilde\nu_1)}
 \nonumber\\
 &\times
 \Gamma\left[
 \frac52+2n_{123}+\I\dd_3\widetilde\nu_2,
 \frac52+2n_4+\I\dd_4\widetilde\nu_2
 \right]
 \nonumber\\
 &\times
 \sin^2\left[
 \frac\pi2\left(\frac52+\I\dd_3\widetilde\nu_2\right)
 \right]
 \left(\frac{r_1}{2}\right)^{2n_{123}
              +\I\dd_3\widetilde\nu_2}
 \left(\frac{r_2}{2}\right)^{2n_4
              +\I\dd_4\widetilde\nu_2},
 \label{eq:tree-nf-propagator-series}
 \\
 \mathcal K_{1|2}^{[\mathrm{FN}]}(r_1,r_2)
 ={}&
 \frac{\I\sinh(\pi\widetilde\nu_2)}{2\pi^2}
 \sum_{\dd_1,\dd_2,\dd_3=\pm1}
 \sum_{n_1,n_2,n_3,n_4=0}^{\infty}
 \dd_3
 \frac{(-1)^{n_{1234}}}{n_1!n_2!n_3!n_4!}
 \nonumber\\
 &\times
 \Gamma\left[
 -n_1-\I\dd_1\widetilde\nu_1,
 -n_2-\I\dd_2\widetilde\nu_1,
 -n_3-\I\dd_3\widetilde\nu_2,
 -n_4+\I\dd_3\widetilde\nu_2
 \right]
 \nonumber\\
 &\times
 \frac{\frac32+2n_2+\I\dd_2\widetilde\nu_1}
 {2n_{23}+\I(\dd_2\widetilde\nu_1+
                  \dd_3\widetilde\nu_2)}
 \nonumber\\
 &\times
 \Gamma\left[
 \frac52+2n_1+\I\dd_1\widetilde\nu_1,
 \frac52+2n_{234}+\I\dd_2\widetilde\nu_1
 \right]
 \nonumber\\
 &\times
 \sin^2\left[
 \frac\pi2\left(\frac52+\I\dd_1\widetilde\nu_1\right)
 \right]
 \left(\frac{r_1}{2}\right)^{2n_1
              +\I\dd_1\widetilde\nu_1}
 \left(\frac{r_2}{2}\right)^{2n_{234}
              +\I\dd_2\widetilde\nu_1}.
 \label{eq:tree-fn-propagator-series}
\end{align}
The denominators in \eqref{eq:tree-nf-propagator-series} and \eqref{eq:tree-fn-propagator-series} are the corresponding values of \(A_M\).  The squared sine factors include the complete SK sums.  These formulas are understood meromorphically at pole collisions.

For the ordered \([\mathrm{NN}]\) kernel, the branch conditions \eqref{eq:nested-branch-selection} cancel the mass-dependent parts of \(s_{1234}\).  Taking the propagator residues in \eqref{eq:tree-pmb-nn} and using the resummed time kernel \eqref{eq:time-nn} gives
\begin{align}
 \mathcal K_{1|2}^{[\mathrm{NN}]}(r_1,r_2)
 ={}&
 \frac{\sinh(\pi\widetilde\nu_1)
       \sinh(\pi\widetilde\nu_2)}{2\pi^2}
 \left(\frac{r_1}{r_2}\right)^{5/2}
 \sum_{\dd_1,\dd_3=\pm1}
 \dd_1\dd_3
 \sum_{n_1,n_2,n_3,n_4=0}^{\infty}
 \nonumber\\
 &\times
 \frac{(-1)^{n_{1234}}}{n_1!n_2!n_3!n_4!}
 \nonumber\\
 &\times
 \Gamma\left[
 -n_1-\I\dd_1\widetilde\nu_1,
 -n_2+\I\dd_1\widetilde\nu_1,
 -n_3-\I\dd_3\widetilde\nu_2,
 -n_4+\I\dd_3\widetilde\nu_2
 \right]
 \nonumber\\
 &\times
 \left(\I\dd_1\widetilde\nu_1-2n_2-\frac32\right)
 \left(\frac{r_1}{2}\right)^{2n_{1234}}
 \nonumber\\
 &\times
 {}_3\mathcal F_2\left[
 \begin{matrix}
 5+2n_{1234},
 \frac52+2n_4-\I\dd_3\widetilde\nu_2,
 \frac52+2n_{234}-\I\dd_1\widetilde\nu_1\\
 \frac72+2n_4-\I\dd_3\widetilde\nu_2,
 \frac72+2n_{234}-\I\dd_1\widetilde\nu_1
 \end{matrix}
 \middle|
 -\frac{r_1}{r_2}
 \right].
 \label{eq:tree-nn-background}
\end{align}
Since the dressed hypergeometric factor is analytic in its argument for \(\lvert r_1/r_2\rvert<1\), the factor multiplying the explicit \((r_1/r_2)^{5/2}\) prefactor is analytic in \(r_1/r_2\) and \(r_1^2\). Restoring the correlator normalization converts the half-integer seed prefactor into \(k_{12}^{-5}\).  The corresponding physical contribution is therefore analytic in both \(k_s^2\) and \(r_1/r_2\).

\paragraph{The fully factorized contribution}

The zero-energy integral in \eqref{eq:time-ff} formally sets \(s_3=-s_2\) when \(\delta\to0\).  At this value the two pole families have no nonzero separation strip, so the central integral is defined with the dimensionless regulator \eqref{eq:tree-middle-time-regulator}.  We take \(\delta\to0\) only after applying Barnes' first lemma \eqref{BarnesLemma}.  Including the Mellin-delta Jacobian fixed by \eqref{eq:mellin-delta-normalization}, define
\begin{align}
 \mathcal B_{12}(\delta)
 \equiv{}&
 \frac12
 \int_{-\I\infty}^{+\I\infty}\frac{\D s_2}{2\pi\I}
 \left(2s_2-\frac32\right)
 \mathcal G_{\widetilde\nu_1}(s_2)
 \mathcal G_{\widetilde\nu_2}\left(\frac\delta2-s_2\right).
 \label{eq:central-barnes}
\end{align}
We decompose the derivative factor as
\begin{align}
 \frac12\left(2s_2-\frac32\right)
 =
 \left(s_2-\frac{\I\widetilde\nu_1}{2}\right)
 +\left(\frac{\I\widetilde\nu_1}{2}-\frac34\right).
 \label{eq:central-barnes-derivative-split}
\end{align}
The second term leaves all four Gamma arguments unshifted. \eqref{BarnesLemma} gives this term a factor \(1/\Gamma(\delta)=\mathcal O(\delta)\), so it vanishes as \(\delta\to0\). In contrast,
\begin{align}
 \left(s_2-\frac{\I\widetilde\nu_1}{2}\right)
 \Gamma\left(s_2-\frac{\I\widetilde\nu_1}{2}\right)
 =
 \Gamma\left(s_2-
 \frac{\I\widetilde\nu_1}{2}+1\right),
 \label{eq:central-barnes-gamma-shift}
\end{align}
and the corresponding Barnes denominator is \(\Gamma(1+\delta)\).  This term therefore has the finite limit
\begin{align}
 \lim_{\delta\to0}\mathcal B_{12}(\delta)
 ={}&
 \Gamma\left[
 1-\frac{\I}{2}(\widetilde\nu_1+\widetilde\nu_2),
 \frac{\I}{2}(\widetilde\nu_1+\widetilde\nu_2),
 1-\frac{\I}{2}(\widetilde\nu_1-\widetilde\nu_2),
 \frac{\I}{2}(\widetilde\nu_1-\widetilde\nu_2)
 \right]
 \nonumber\\
 ={}&
 -\frac{\pi^2}{
 \sinh\left[\frac\pi2(\widetilde\nu_1+\widetilde\nu_2)\right]
 \sinh\left[\frac\pi2(\widetilde\nu_1-\widetilde\nu_2)\right]}.
 \label{eq:central-barnes-result}
\end{align}

In the residue series obtained from the complete eight-branch SK sum, the two endpoint levels are \(n_1,n_4\in\mathbb N_0\).  The two sine factors together supply \((-1)^{n_1+n_4}\), while the cosine supplies the same sign.  Their product is therefore independent of \(n_1,n_4\). The two endpoint sums separate and can be resummed in closed form, giving
\begin{align}
\mathcal K_{1|2}^{[\mathrm{FF}]}(r_1,r_2)
 ={}&
 -\frac{1}{
 2\sinh\left[\frac\pi2(\widetilde\nu_1+\widetilde\nu_2)\right]
 \sinh\left[\frac\pi2(\widetilde\nu_1-\widetilde\nu_2)\right]}
 \sum_{\dd_1,\dd_4=\pm1}
 \nonumber\\
 &\times
 \Gamma\left[
 -\I\dd_1\widetilde\nu_1,
 \frac52+\I\dd_1\widetilde\nu_1,
 -\I\dd_4\widetilde\nu_2,
 \frac52+\I\dd_4\widetilde\nu_2
 \right]
 \nonumber\\
 &\times
 \left(\frac{r_1}{2}\right)^{\I\dd_1\widetilde\nu_1}
 \left(\frac{r_2}{2}\right)^{\I\dd_4\widetilde\nu_2}
 \nonumber\\
 &\times
 \sin\left[\frac\pi2\left(
 \frac52+\I\dd_1\widetilde\nu_1\right)\right]
 \sin\left[\frac\pi2\left(
 \frac52+\I\dd_4\widetilde\nu_2\right)\right]
 \nonumber\\
 &\times
 \cos\left[\frac\pi2\left(
 5
 +\I\dd_1\widetilde\nu_1
 +\I\dd_4\widetilde\nu_2\right)\right]
 \nonumber\\
 &\times
 {}_2\mathrm F_1\left[\begin{matrix}
 \frac54+\frac{\I\dd_1\widetilde\nu_1}{2},
 \frac74+\frac{\I\dd_1\widetilde\nu_1}{2}\\
 1+\I\dd_1\widetilde\nu_1
 \end{matrix}\middle|r_1^2\right]
 {}_2\mathrm F_1\left[\begin{matrix}
 \frac54+\frac{\I\dd_4\widetilde\nu_2}{2},
 \frac74+\frac{\I\dd_4\widetilde\nu_2}{2}\\
 1+\I\dd_4\widetilde\nu_2
 \end{matrix}\middle|r_2^2\right].
 \label{eq:fully-factorized-series}
\end{align}
This closed form holds for \(0<r_1,r_2<1\) and is meromorphic in \(\widetilde\nu_{1,2}\).  It is continued to the physical masses below only after all general-mass contributions have been summed.

\paragraph{The opposite orientation}

For \(\mathcal K_{2|1}\), the Mellin variables remain positional: \((s_1,s_2)\) labels the left line with mass \(\widetilde\nu_2\), and \((s_3,s_4)\) the right line with mass \(\widetilde\nu_1\).  The right-line derivative acts on its first time argument and supplies \(2s_3-3/2\). Thus the four PMB formulas \eqref{eq:tree-pmb-ff}--\eqref{eq:tree-pmb-nn} give the corresponding \(2|1\) contributions after the replacement
\begin{align}
 &\left(2s_2-\frac32\right)
 \prod_{i=1}^{2}\mathcal G_{\widetilde\nu_1}(s_i)
 \prod_{j=3}^{4}\mathcal G_{\widetilde\nu_2}(s_j)
 \nonumber\\
 &\hspace{1.5cm}\longrightarrow
 \left(2s_3-\frac32\right)
 \prod_{i=1}^{2}\mathcal G_{\widetilde\nu_2}(s_i)
 \prod_{j=3}^{4}\mathcal G_{\widetilde\nu_1}(s_j).
 \label{eq:tree-pmb-21-replacement}
\end{align}
The positional SK phases, the four sector prefactors, the time exponents in \eqref{eq:tree-time-exponents}, and all four time kernels remain unchanged; the replacement \eqref{eq:tree-pmb-21-replacement} applies both to the four PMB integrands and to their left-pole residue sums.

In the fully factorized sector, the Mellin delta sets \(s_2=\delta/2-s_3\).  After relabeling the integration variable, the central integral is exactly \(\mathcal B_{12}(\delta)\) in \eqref{eq:central-barnes}. Its limit is still \eqref{eq:central-barnes-result}, because the derivative continues to act on the line with mass \(\widetilde\nu_1\). The same cancellation of the level-dependent trigonometric signs and endpoint resummation give
\begin{align}
 \mathcal K_{2|1}^{[\mathrm{FF}]}(r_1,r_2)
 ={}&
 -\frac{1}{
 2\sinh\left[\frac\pi2(\widetilde\nu_1+\widetilde\nu_2)\right]
 \sinh\left[\frac\pi2(\widetilde\nu_1-\widetilde\nu_2)\right]}
 \sum_{\dd_1,\dd_4=\pm1}
 \nonumber\\
 &\times
 \Gamma\left[
 -\I\dd_1\widetilde\nu_2,
 \frac52+\I\dd_1\widetilde\nu_2,
 -\I\dd_4\widetilde\nu_1,
 \frac52+\I\dd_4\widetilde\nu_1
 \right]
 \nonumber\\
 &\times
 \left(\frac{r_1}{2}\right)^{\I\dd_1\widetilde\nu_2}
 \left(\frac{r_2}{2}\right)^{\I\dd_4\widetilde\nu_1}
 \nonumber\\
 &\times
 \sin\left[\frac\pi2\left(
 \frac52+\I\dd_1\widetilde\nu_2\right)\right]
 \sin\left[\frac\pi2\left(
 \frac52+\I\dd_4\widetilde\nu_1\right)\right]
 \nonumber\\
 &\times
 \cos\left[\frac\pi2\left(
 5
 +\I\dd_1\widetilde\nu_2
 +\I\dd_4\widetilde\nu_1\right)\right]
 \nonumber\\
 &\times
 {}_2\mathrm F_1\left[\begin{matrix}
 \frac54+\frac{\I\dd_1\widetilde\nu_2}{2},
 \frac74+\frac{\I\dd_1\widetilde\nu_2}{2}\\
 1+\I\dd_1\widetilde\nu_2
 \end{matrix}\middle|r_1^2\right]
 {}_2\mathrm F_1\left[\begin{matrix}
 \frac54+\frac{\I\dd_4\widetilde\nu_1}{2},
 \frac74+\frac{\I\dd_4\widetilde\nu_1}{2}\\
 1+\I\dd_4\widetilde\nu_1
 \end{matrix}\middle|r_2^2\right].
 \label{eq:fully-factorized-series-21}
\end{align}
For the nested sectors, the branch conditions \eqref{eq:nested-branch-selection} remain positional.  Evaluating the new derivative factor \(2s_3-3/2\) on these poles gives
\begin{align}
 \mathcal K_{2|1}^{[\mathrm{NF}]}(r_1,r_2)
 ={}&
 \frac{\I\sinh(\pi\widetilde\nu_2)}{2\pi^2}
 \sum_{\dd_1,\dd_3,\dd_4=\pm1}
 \sum_{n_1,n_2,n_3,n_4=0}^{\infty}
 \dd_1
 \frac{(-1)^{n_{1234}}}{n_1!n_2!n_3!n_4!}
 \nonumber\\
 &\times
 \Gamma\left[
 -n_1-\I\dd_1\widetilde\nu_2,
 -n_2+\I\dd_1\widetilde\nu_2,
 -n_3-\I\dd_3\widetilde\nu_1,
 -n_4-\I\dd_4\widetilde\nu_1
 \right]
 \nonumber\\
 &\times
 \frac{-2n_3-\frac32-\I\dd_3\widetilde\nu_1}
 {2n_{23}+\I(\dd_3\widetilde\nu_1-
                  \dd_1\widetilde\nu_2)}
 \nonumber\\
 &\times
 \Gamma\left[
 \frac52+2n_{123}+\I\dd_3\widetilde\nu_1,
 \frac52+2n_4+\I\dd_4\widetilde\nu_1
 \right]
 \nonumber\\
 &\times
 \sin^2\left[
 \frac\pi2\left(\frac52+\I\dd_3\widetilde\nu_1\right)
 \right]
 \left(\frac{r_1}{2}\right)^{2n_{123}
              +\I\dd_3\widetilde\nu_1}
 \left(\frac{r_2}{2}\right)^{2n_4
              +\I\dd_4\widetilde\nu_1},
 \label{eq:tree-nf-propagator-series-21}
\end{align}
and
\begin{align}
 \mathcal K_{2|1}^{[\mathrm{FN}]}(r_1,r_2)
 ={}&
 \frac{\I\sinh(\pi\widetilde\nu_1)}{2\pi^2}
 \sum_{\dd_1,\dd_2,\dd_3=\pm1}
 \sum_{n_1,n_2,n_3,n_4=0}^{\infty}
 \dd_3
 \frac{(-1)^{n_{1234}}}{n_1!n_2!n_3!n_4!}
 \nonumber\\
 &\times
 \Gamma\left[
 -n_1-\I\dd_1\widetilde\nu_2,
 -n_2-\I\dd_2\widetilde\nu_2,
 -n_3-\I\dd_3\widetilde\nu_1,
 -n_4+\I\dd_3\widetilde\nu_1
 \right]
 \nonumber\\
 &\times
 \frac{\frac32+2n_3+\I\dd_3\widetilde\nu_1}
 {2n_{23}+\I(\dd_2\widetilde\nu_2+
                  \dd_3\widetilde\nu_1)}
 \nonumber\\
 &\times
 \Gamma\left[
 \frac52+2n_1+\I\dd_1\widetilde\nu_2,
 \frac52+2n_{234}+\I\dd_2\widetilde\nu_2
 \right]
 \nonumber\\
 &\times
 \sin^2\left[
 \frac\pi2\left(\frac52+\I\dd_1\widetilde\nu_2\right)
 \right]
 \left(\frac{r_1}{2}\right)^{2n_1
              +\I\dd_1\widetilde\nu_2}
 \left(\frac{r_2}{2}\right)^{2n_{234}
              +\I\dd_2\widetilde\nu_2}.
 \label{eq:tree-fn-propagator-series-21}
\end{align}
Thus \([\mathrm{NF}]\) carries the two-exponent structures of \(\widetilde\nu_1\), whereas \([\mathrm{FN}]\) carries those of \(\widetilde\nu_2\).

For the ordered fully nested contribution, the two branch conditions cancel the mass-dependent parts of \(s_{1234}\).  Using \eqref{eq:time-nn}, the remaining result is
\begin{align}
 \mathcal K_{2|1}^{[\mathrm{NN}]}(r_1,r_2)
 ={}&
 \frac{\sinh(\pi\widetilde\nu_1)
       \sinh(\pi\widetilde\nu_2)}{2\pi^2}
 \left(\frac{r_1}{r_2}\right)^{5/2}
 \sum_{\dd_1,\dd_3=\pm1}
 \dd_1\dd_3
 \sum_{n_1,n_2,n_3,n_4=0}^{\infty}
 \nonumber\\
 &\times
 \frac{(-1)^{n_{1234}}}{n_1!n_2!n_3!n_4!}
 \nonumber\\
 &\times
 \Gamma\left[
 -n_1-\I\dd_1\widetilde\nu_2,
 -n_2+\I\dd_1\widetilde\nu_2,
 -n_3-\I\dd_3\widetilde\nu_1,
 -n_4+\I\dd_3\widetilde\nu_1
 \right]
 \nonumber\\
 &\times
 \left(-\I\dd_3\widetilde\nu_1-2n_3-\frac32\right)
 \left(\frac{r_1}{2}\right)^{2n_{1234}}
 \nonumber\\
 &\times
 {}_3\mathcal F_2\left[
 \begin{matrix}
 5+2n_{1234},
 \frac52+2n_4-\I\dd_3\widetilde\nu_1,
 \frac52+2n_{234}-\I\dd_1\widetilde\nu_2\\
 \frac72+2n_4-\I\dd_3\widetilde\nu_1,
 \frac72+2n_{234}-\I\dd_1\widetilde\nu_2
 \end{matrix}
 \middle|
 -\frac{r_1}{r_2}
 \right].
 \label{eq:tree-nn-background-21}
\end{align}
The Taylor-expansion argument below \eqref{eq:tree-nn-background} again shows that this contribution is analytic after restoring the correlator normalization. \eqref{eq:fully-factorized-series-21}, \eqref{eq:tree-nf-propagator-series-21}, \eqref{eq:tree-fn-propagator-series-21}, and \eqref{eq:tree-nn-background-21} give \(\mathcal K_{2|1}\) directly in the same configuration \(r_1<r_2\), without exchanging a small-\(r_1/r_2\) series term by term.

\subsection{Physical correlator and hierarchical squeezed limit}

Following \eqref{eq:propagator-mass-continuation}, we continue the two general mass parameters according to
\begin{align}
 \widetilde\nu_1\longrightarrow-\I\nu_\pi,
 \qquad
 \widetilde\nu_2\longrightarrow\widetilde\nu_\sigma,
 \qquad
 0<\nu_\pi<\frac32,
 \qquad
 \widetilde\nu_\sigma>0.
 \label{eq:tree-physical-continuation}
\end{align}
This continuation is applied to the complete SK-summed expressions in the preceding subsection by continuous contour deformation.  The two orientations become \(\mathcal K_{1|2}\longrightarrow\mathcal K_{\pi|\sigma}\) and \(\mathcal K_{2|1}\longrightarrow\mathcal K_{\sigma|\pi}\), where the derivative follows the complementary field \(\pi\) in both cases. Together with \eqref{T2K2ptMixed} and \eqref{TreeSeed}, these continuations give the physical correlator.  The two oriented seeds entering it are obtained directly from \eqref{eq:fully-factorized-series}, \eqref{eq:tree-nf-propagator-series}, \eqref{eq:tree-fn-propagator-series}, and \eqref{eq:tree-nn-background}, together with their opposite-orientation counterparts \eqref{eq:fully-factorized-series-21}, \eqref{eq:tree-nf-propagator-series-21}, \eqref{eq:tree-fn-propagator-series-21}, and \eqref{eq:tree-nn-background-21}, after making the physical continuation above.

For each orientation, the superscripts \([\mathrm{FF}]\), \([\mathrm{NF}]\), \([\mathrm{FN}]\), and \([\mathrm{NN}]\) denote the physical continuations of the four sector results in the preceding subsection.  Using the physical decomposition \eqref{eq:main-tree-component-decomposition}, their exact definitions are
\begin{align}
 \mathcal K_{\mathrm{mix}}
 &\equiv
 \mathcal K_{\pi|\sigma}^{[\mathrm{FF}]}
 +
 \mathcal K_{\sigma|\pi}^{[\mathrm{FF}]},
 \label{Kmix}
 \\
 \mathcal K_{\mathrm{sca}}
 &\equiv
 \mathcal K_{\pi|\sigma}^{[\mathrm{FN}]}
 +
 \mathcal K_{\sigma|\pi}^{[\mathrm{NF}]},
 \label{Ksca}
 \\
 \mathcal K_{\mathrm{osc}}
 &\equiv
 \mathcal K_{\pi|\sigma}^{[\mathrm{NF}]}
 +
 \mathcal K_{\sigma|\pi}^{[\mathrm{FN}]},
 \label{Kosc}
 \\
 \mathcal K_{\mathrm{bg}}
 &\equiv
 \mathcal K_{\pi|\sigma}^{[\mathrm{NN}]}
 +
 \mathcal K_{\sigma|\pi}^{[\mathrm{NN}]}.
 \label{Kbg}
\end{align}
The abbreviation ``\(\mathrm{sca}\)'' denotes the nonoscillatory complementary-scaling component, whereas ``\(\mathrm{osc}\)'' denotes the principal-series oscillatory component.

\paragraph{Leading behavior in the hierarchical squeezed limit}

For fixed masses and \(0<r_1\ll r_2\ll1\), taking the leading residue levels in the four physical components gives
\begin{align}
 \lim_{r_1\ll r_2\ll1}
 \mathcal K_{\mathrm{mix}}(r_1,r_2)
 ={}&
 \sum_{\bb=\pm1}
 C^{\mathrm{mix}}_{\bb}
 (\nu_\pi,\widetilde\nu_\sigma)
 r_1^{-\nu_\pi}
 r_2^{\I\bb\widetilde\nu_\sigma},
 \label{eq:leading-mixed-component}
 \\
 \lim_{r_1\ll r_2\ll1}
 \mathcal K_{\mathrm{sca}}(r_1,r_2)
 ={}&
 C^{\mathrm{sca}}(\nu_\pi,\widetilde\nu_\sigma)
 (r_1r_2)^{-\nu_\pi},
 \label{eq:leading-scaling-component}
 \\
 \lim_{r_1\ll r_2\ll1}
 \mathcal K_{\mathrm{osc}}(r_1,r_2)
 ={}&
 \sum_{\aa,\bb=\pm1}
 C^{\mathrm{osc}}_{\aa\bb}
 (\nu_\pi,\widetilde\nu_\sigma)
 r_1^{\I\aa\widetilde\nu_\sigma}
 r_2^{\I\bb\widetilde\nu_\sigma},
 \label{eq:leading-oscillatory-component}
 \\
 \lim_{r_1\ll r_2\ll1}
 \mathcal K_{\mathrm{bg}}(r_1,r_2)
 ={}&
 C^{\mathrm{bg}}(\nu_\pi,\widetilde\nu_\sigma)
 \left(\frac{r_1}{r_2}\right)^{5/2}.
 \label{eq:leading-background-component}
\end{align}
Here all coefficients include both orientations in \eqref{Kmix}--\eqref{Kbg}, except that the strict hierarchical limit in \eqref{eq:leading-mixed-component} retains only the dominant \(\pi|\sigma\) orientation.  They are
\begin{align}
 C^{\mathrm{mix}}_{\bb}
 (\nu_\pi,\widetilde\nu_\sigma)
 ={}&
 2^{\nu_\pi-\I\bb\widetilde\nu_\sigma-1}
 \Gamma\left[
 \nu_\pi,\frac52-\nu_\pi,
 -\I\bb\widetilde\nu_\sigma,
 \frac52+\I\bb\widetilde\nu_\sigma
 \right]
 \nonumber\\
 &\times
 \frac{
 \sin\left[\frac\pi2\left(\frac52-\nu_\pi\right)\right]
 \sin\left[\frac\pi2\left(
 \frac52+\I\bb\widetilde\nu_\sigma\right)\right]
 }{
 \sin\left[\frac\pi2\left(
 \nu_\pi+\I\bb\widetilde\nu_\sigma\right)\right]
 },
 \label{eq:leading-mixed-coefficient}
 \\
 C^{\mathrm{sca}}(\nu_\pi,\widetilde\nu_\sigma)
 ={}&
 \frac{2^{2\nu_\pi}}{\pi}
 \frac{
 \left(\frac32-\nu_\pi\right)
 \left[1-\sin(\pi\nu_\pi)\right]
 }{
 \nu_\pi^2+\widetilde\nu_\sigma^2
 }
 \Gamma\left[
 \nu_\pi,\nu_\pi,
 \frac52-\nu_\pi,\frac52-\nu_\pi
 \right],
 \label{eq:leading-scaling-coefficient}
 \\
 C^{\mathrm{osc}}_{\aa\bb}
 (\nu_\pi,\widetilde\nu_\sigma)
 ={}&
 -\frac{
 2^{-\I(\aa+\bb)\widetilde\nu_\sigma}
 \left[3-\I(\aa+\bb)\widetilde\nu_\sigma\right]
 }{
 2\pi\left(\nu_\pi^2+\widetilde\nu_\sigma^2\right)
 }
 \nonumber\\
 &\times
 \left[1+\I\aa\sinh(\pi\widetilde\nu_\sigma)\right]
 \nonumber\\
 &\times
 \Gamma\left[
 -\I\aa\widetilde\nu_\sigma,
 -\I\bb\widetilde\nu_\sigma,
 \frac52+\I\aa\widetilde\nu_\sigma,
 \frac52+\I\bb\widetilde\nu_\sigma
 \right],
 \label{eq:leading-oscillatory-coefficient}
 \\
 C^{\mathrm{bg}}(\nu_\pi,\widetilde\nu_\sigma)
 ={}&
 \frac{144}{
 \left(\frac{25}{4}-\nu_\pi^2\right)
 \left(\frac{25}{4}+\widetilde\nu_\sigma^2\right)}.
 \label{eq:leading-background-coefficient}
\end{align}
The coefficients in \eqref{eq:leading-mixed-coefficient}-- \eqref{eq:leading-background-coefficient} follow by setting the relevant residue levels to zero in the series of the preceding subsection.  For the background, the dressed hypergeometric function at the origin retains the Gamma ratio in its definition \eqref{eq:dressed-hypergeometric}; including this factor gives \eqref{eq:leading-background-coefficient}.

Restoring the momentum prefactor in \eqref{T2K2ptMixed} gives the schematic hierarchical scaling
\begin{equation}
\label{eq_4ptMixbehavior}
    \lim_{r_1\ll r_2\ll 1} \mathcal T_s \sim \frac{1}{k_1k_2k_3k_4k_s^5}\bigg\{ (r_1r_2)^{5/2} \Big[ (r_1r_2)^{-\nu_\pi} + r_1^{-\nu_\pi}r_2^{\pm\I\widetilde\nu_\sigma}
     + r_1^{\pm\I\wt\nu_\sigma}r_2^{-\nu_\pi}
     + \cdots\Big] + r_1^5 \bigg\}.
\end{equation}
Here the couplings and mass-dependent coefficients are omitted.

The leading \(\sigma|\pi\) contribution to the mixed component scales as \(r_1^{\I\bb\widetilde\nu_\sigma}r_2^{-\nu_\pi}\) and is suppressed relative to \eqref{eq:leading-mixed-component} by \((r_1/r_2)^{\nu_\pi}\).  The reflected \(\pi|\sigma\) branch \(r_1^{+\nu_\pi}r_2^{\I\bb\widetilde\nu_\sigma}\) is suppressed by \(r_1^{2\nu_\pi}\), while its orientation-reversed counterpart \(r_1^{\I\bb\widetilde\nu_\sigma}r_2^{+\nu_\pi}\) is suppressed by \((r_1r_2)^{\nu_\pi}\).  Higher signal descendants carry additional even powers of \(r_1\) or \(r_2\).

Notice that \(\nu_\pi=1/2\) is a special case.  At this point, the light-endpoint sine in \eqref{eq:leading-mixed-coefficient} vanishes at every descendant level, and \eqref{eq:leading-scaling-coefficient} also vanishes.  The leading growing-\(\pi\) branches of the mixed and complementary-scaling components are therefore absent.  The first surviving complementary-dependent descendant scales as \((r_1/r_2)^{1/2}\), while the principal-series oscillatory component remains of order unity.  Hence the ordering stated above does not apply at this point.  At \(\nu_\pi\to0\) or \(\widetilde\nu_\sigma\to0\), pole branches coalesce and must be combined before taking the limit.  More generally, coincident pole families are defined with a mass regulator and can generate logarithmic terms.

\paragraph{Large-mass limit of the mixed component}

Applying the Gamma-function asymptotics \eqref{GammaLimit} to \eqref{eq:leading-mixed-coefficient}, for fixed \(\nu_\pi\) and \(\widetilde\nu_\sigma\gg1\), gives the power-form result
\begin{align}
 \lim_{r_1\ll r_2\ll1}
 \mathcal K_{\mathrm{mix}}(r_1,r_2)
 ={}&
 \pi
 \Gamma\left[\nu_\pi,\frac52-\nu_\pi\right]
 \sin\left[\frac\pi2\left(\frac52-\nu_\pi\right)\right]
 \widetilde\nu_\sigma^{3/2}
 \E^{-\pi\widetilde\nu_\sigma}
 \left(\frac{r_1}2\right)^{-\nu_\pi}
 \nonumber\\
 &\times
 \sum_{\bb=\pm1}
 \E^{\I\bb\pi\nu_\pi/2}
 \left(\frac{r_2}{2}\right)^{
 \I\bb\widetilde\nu_\sigma}
 \left[1+\mathcal O(\widetilde\nu_\sigma^{-1})\right].
 \label{eq:leading-mixed-large-mass}
\end{align}

Combining the two conjugate branches gives the real leading-order expression \eqref{K12Nu} in the main text.

   \section{Loop Seed Integral With Two General Principal-Series Scalars}\label{CalLoop}
We evaluate \(\mathcal J_{\wt\nu_1,\wt\nu_2}(r)\) defined in \eqref{LoopSeedIntDef}, with \(r\) defined in \eqref{BinJ}, initially taking both mass parameters on the principal-series line according to \eqref{PriModeF}.  The special-function notation follows App.~\ref{SpecialFunctions}. The seed definition \eqref{LoopSeedIntDef} contains no diagrammatic symmetry factor.  For an identical-field loop, the factor \(1/2\) is supplied separately in \eqref{eq_Bdef}; it is absent for two distinct internal fields.

\subsection{Spectral representation and renormalization}

We evaluate \eqref{LoopSeedIntDef} in dimensional regularization using the spectral (K\"all\'{e}n--Lehmann) representation \cite{Marolf:2010zp,Sleight:2021plv,DiPietro:2021sjt,Loparco:2023rug,Xianyu:2022jwk,Zhang:2025nzd}. In \(d\) spatial dimensions, the dimensionally regulated identity is
\begin{align}\label{eq_specD}
\int \FR{\D^d\bm q}{(2\pi)^d}\,
D^{(\wt\nu_1)}_{\mathsf a\mathsf b}(|\bm q|;\tau_1,\tau_2)
D^{(\wt\nu_2)}_{\mathsf a\mathsf b}(|\bm k-\bm q|;\tau_1,\tau_2)
= \int_{-\infty}^{+\infty} \D\wt\nu'\,\FR{\wt\nu'}{\pi\I}
\rho^d_{\wt\nu_1,\wt\nu_2}(\wt\nu')
D^{(\wt\nu')}_{\mathsf a\mathsf b}(k;\tau_1,\tau_2),
\end{align}
where \(\rho^d_{\widetilde\nu_1,\widetilde\nu_2}\) is the spectral density and \(D\) denotes the dimensionally continued \((d+1)\)-dimensional propagator, with \(\widetilde\nu_i\) held fixed during the dimensional continuation. The identity is initially used for two principal-series parameters and expresses the bubble as a one-dimensional superposition of tree exchanges with effective mass parameter \(\widetilde\nu'\). The spectral function in \((d+1)\)-dimensional dS space is given by~\cite{Marolf:2010zp}. We use the dressed hypergeometric convention \eqref{eq:dressed-hypergeometric}:
   \begin{align}\label{eq:loop-spectral-function}
{}&  \rho^d_{\wt\nu_1,\wt\nu_2}(\wt\nu') =\bigg\{
            \frac{-1}{2(4\pi)^{(d+1)/2}}
            \frac{\cos\left[\pi\left(-\frac{d}{2}+\I\wt\nu_1\right)\right]}{\sin\left(\pi\I\wt\nu_1\right)}
            \Gamma\left[\begin{matrix}
            \frac{3-d}2,
            \frac{d}{2}-\I\wt\nu_1
            \\
            1-\frac d2-\I\wt\nu_1
            \end{matrix}
            \right]
        \nonumber\\
        &\times
            {}_{7}\mathcal F_{6}\left[
            \begin{matrix}
            1-\frac{d}{2}+\I\wt\nu'-\I\wt\nu_1,
            \frac{3-d/2+\I\wt\nu'-\I\wt\nu_1}{2},
            1-\frac{d}{2},
            1-\frac{d}{2}-\I\wt\nu_1,
            1-\frac{d}{2}+\I\wt\nu',
            \frac{d/2+\I\wt\nu'-\I\wt\nu_1-\I\wt\nu_2}{2},
            \frac{d/2+\I\wt\nu'-\I\wt\nu_1+\I\wt\nu_2}{2}\\
            \frac{1-d/2+\I\wt\nu'-\I\wt\nu_1}{2},
            1+\I\wt\nu'-\I\wt\nu_1,
            1+\I\wt\nu',
            1-\I\wt\nu_1,
            \frac{4-3d/2+\I\wt\nu'-\I\wt\nu_1-\I\wt\nu_2}{2},
            \frac{4-3d/2+\I\wt\nu'-\I\wt\nu_1+\I\wt\nu_2}{2}
            \end{matrix}
            \middle|1\right]\nonumber\\
            {}&+\left(\wt\nu_1 \to -\wt\nu_1 \right)
            \bigg\}
            +\left(\wt\nu_1\leftrightarrow\wt\nu_2\right).
\end{align}
The pole of \(\Gamma[(3-d)/2]\) in the first line gives the mass- and \(\wt\nu'\)-independent UV divergence \cite{Marolf:2010zp,Xianyu:2022jwk,Zhang:2025nzd}
\begin{align}
\rho^d_{\wt\nu_1,\wt\nu_2}(\wt\nu')
=
\frac{1}{8\pi^2(d-3)}+\text{finite},
\qquad d\to3,
\label{eq:loop-spectral-divergence}
\end{align}
which agrees with the flat-space UV divergence.  Following \cite{Xianyu:2022jwk}, we define the \(\overline{\mathrm{MS}}\)-subtracted spectral function in \(3+1\) dimensions by
\begin{align}
\wh\rho_{\wt\nu_1,\wt\nu_2}(\wt\nu')
\equiv{}&
\lim_{d\to 3}\left[
\rho^d_{\wt\nu_1,\wt\nu_2}(\wt\nu')
-\frac{1}{(4\pi)^2}\left(
\frac{2}{d-3}+\gamma_E-\log 4\pi
\right)\right].
\label{eq:loop-renormalized-spectral-function}
\end{align}
Here \(\gamma_E\) is the Euler--Mascheroni constant.  The renormalization scale enters the background through the explicit combination \(\wh\rho-(4\pi)^{-2}\log\mu_R^2\); the antisymmetric spectral function used below is independent of this subtraction.  Combining \eqref{LoopSeedIntDef}, the dimensionally continued version of \eqref{eq_specD}, and the tree-seed definition \eqref{eq_Idef}, we obtain
\begin{align}\label{eq:loop-spectral-integral}
\mathcal J_{\wt\nu_1,\wt\nu_2}(r)
=&~\int_{-\infty}^{+\infty} \D\wt\nu'\,\FR{\wt\nu'}{\pi\I}
\left[
\wh\rho_{\wt\nu_1,\wt\nu_2}(\wt\nu')
-\frac{1}{(4\pi)^2}\log\mu_R^2
\right]
\mathcal I_{\wt\nu'}^{0,-2}(r,1).
\end{align}

\subsection{Pole structure and residue evaluation}

We evaluate the spectral integral \eqref{eq:loop-spectral-integral} by closing its contour in the complex \(\wt\nu'\) plane.

The subtraction in \eqref{eq:loop-renormalized-spectral-function} cancels from the antisymmetric spectral function \cite{Zhang:2025nzd},
\begin{align}
\Pi_{\wt\nu_1,\wt\nu_2}(\wt\nu')
\equiv{}&
\wh\rho_{\wt\nu_1,\wt\nu_2}(\wt\nu')
-\wh\rho_{\wt\nu_1,\wt\nu_2}(-\wt\nu')
\nonumber\\
={}&
\frac{\I\sinh(\pi\wt\nu')}
{8\pi^4\Gamma[3/2-\I\wt\nu',3/2+\I\wt\nu']}
\prod_{\dd_1,\dd_2,\dd_3=\pm1}
\Gamma\!\left(
\frac{3/2+\I\dd_1\wt\nu_1
+\I\dd_2\wt\nu_2+\I\dd_3\wt\nu'}{2}
\right).
\label{eq:loop-antisymmetric-spectral-function}
\end{align}
The product in \eqref{eq:loop-antisymmetric-spectral-function} is over all eight combinations of the three independent signs. For real principal-series parameters \(\wt\nu_1\) and \(\wt\nu_2\), its upper-half-plane poles, and hence those of \(\wh\rho\), are
\begin{align}
\wt\nu'=\mu_n^{\dd_1\dd_2}
&\equiv
\I\left(\frac32+2n\right)
+\dd_1\wt\nu_1+\dd_2\wt\nu_2,
&
\dd_1,\dd_2&=\pm1,
\qquad n\in\mathbb N_0.
\label{eq:loop-spectral-poles}
\end{align}

\paragraph{Folded tree seed and contour prescription}

The three terms in the folded seed \eqref{eq_foldedI} have different decay properties in the complex \(\wt\nu'\) plane.  Rewriting its dressed \({}_2\mathcal F_1\) terms using \eqref{eq:dressed-hypergeometric}, we separate the two signal branches and the background as
\begin{align}
\wt\nu'\mathcal I^{0,-2}_{\wt\nu'}(r,1)
={}&
\wt{\mathcal I}_{\wt\nu'}^{(1)}(r)
+\wt{\mathcal I}_{\wt\nu'}^{(2)}(r)
+\wt{\mathcal I}_{\wt\nu'}^{(3)}(r),
\label{eq:loop-folded-integrand-decomposition}
\\
\wt{\mathcal I}_{\wt\nu'}^{(1)}(r)
={}&
-\wt\nu'
\frac{1-\I\sinh(\pi\wt\nu')}
{2^{1/2-\I\wt\nu'}\pi^{1/2}}
\left(\frac{r}{1+r}\right)^{5/2-\I\wt\nu'}
\nonumber\\
&\times
\Gamma\left[
\frac52-\I\wt\nu',\I\wt\nu',
\frac12-\I\wt\nu',\frac12+\I\wt\nu'
\right]
{}_2\mathrm F_1\left[
\begin{matrix}
\frac12-\I\wt\nu',\frac52-\I\wt\nu'\\
1-2\I\wt\nu'
\end{matrix}
\middle|\frac{2r}{1+r}
\right],
\label{eq:loop-folded-signal-branch-1}
\\
\wt{\mathcal I}_{\wt\nu'}^{(2)}(r)
={}&
-\wt\nu'
\frac{1+\I\sinh(\pi\wt\nu')}
{2^{1/2+\I\wt\nu'}\pi^{1/2}}
\left(\frac{r}{1+r}\right)^{5/2+\I\wt\nu'}
\nonumber\\
&\times
\Gamma\left[
\frac52+\I\wt\nu',-\I\wt\nu',
\frac12-\I\wt\nu',\frac12+\I\wt\nu'
\right]
{}_2\mathrm F_1\left[
\begin{matrix}
\frac12+\I\wt\nu',\frac52+\I\wt\nu'\\
1+2\I\wt\nu'
\end{matrix}
\middle|\frac{2r}{1+r}
\right],
\label{eq:loop-folded-signal-branch-2}
\\
\wt{\mathcal I}_{\wt\nu'}^{(3)}(r)
={}&
-\frac{4\wt\nu'}{(\wt\nu')^2+1/4}
\left(\frac{r}{1+r}\right)^3
{}_3\mathrm F_2\left[
\begin{matrix}
1,1,3\\
\frac32-\I\wt\nu',\frac32+\I\wt\nu'
\end{matrix}
\middle|\frac{2r}{1+r}
\right].
\label{eq:loop-folded-background}
\end{align}
For \(0<r<1\), the Gamma asymptotics \eqref{GammaLimit} require the contour of \(\wt{\mathcal I}^{(1)}\) to be closed in the upper half-plane and that of \(\wt{\mathcal I}^{(2)}\) in the lower half-plane.  Their candidate poles can be written directly as the half-integer grids
\begin{align}
\wt{\mathcal I}^{(1)}:\quad
\wt\nu'&=\frac{\I(m+1)}2,
&
\wt{\mathcal I}^{(2)}:\quad
\wt\nu'&=-\frac{\I(m+1)}2,
&
m&\in\mathbb N_0.
\label{eq:loop-signal-branch-poles}
\end{align}
The points \(m\equiv2\pmod 4\) are removable and are retained in \eqref{eq:loop-signal-branch-poles} as zero-residue terms.  The background contour may be closed in either half-plane \cite{Zhang:2025nzd}; choosing the lower half-plane avoids the spectral poles and encloses
\begin{align}
\wt{\mathcal I}^{(3)}:\quad
\wt\nu'=-\I\left(n+\frac12\right),
\qquad n\in\mathbb N_0.
\label{eq:loop-background-poles}
\end{align}
The latter sequence includes the pole of the explicit rational factor at \(n=0\) and the lower-parameter poles of the ordinary \({}_3\mathrm F_2\), defined in \eqref{eq:ordinary-hypergeometric}, for \(n\geq1\).

The two signal branches obey
\begin{align}
\wt{\mathcal I}^{(2)}_{-\wt\nu'}(r)
={}&-\wt{\mathcal I}^{(1)}_{\wt\nu'}(r),
&
\operatorname*{Res}_{\wt\nu'=-\I(m+1)/2}
\wt{\mathcal I}^{(2)}_{\wt\nu'}(r)
={}&
\operatorname*{Res}_{\wt\nu'=\I(m+1)/2}
\wt{\mathcal I}^{(1)}_{\wt\nu'}(r).
\label{eq:loop-folded-branch-reflection}
\end{align}
Because the original contour runs from \(-\infty\) to \(+\infty\), upper- and lower-half-plane closures carry residue factors \(+2\) and \(-2\), respectively.  The complete residue sum is therefore
\begin{align}
\mathcal J_{\wt\nu_1,\wt\nu_2}(r)
={}&
\sum_{n=0}^{\infty}
\sum_{\dd_1,\dd_2=\pm1}
2\,\wt{\mathcal I}^{(1)}_{\mu_n^{\dd_1\dd_2}}(r)
\operatorname*{Res}_{\wt\nu'=\mu_n^{\dd_1\dd_2}}
\Pi_{\wt\nu_1,\wt\nu_2}(\wt\nu')
\nonumber\\
&+
\sum_{m=0}^{\infty}
2\,\Pi_{\wt\nu_1,\wt\nu_2}\left(\frac{\I(m+1)}2\right)
\operatorname*{Res}_{\wt\nu'=\I(m+1)/2}
\wt{\mathcal I}^{(1)}_{\wt\nu'}(r)
\nonumber\\
&-
\sum_{n=0}^{\infty}
2\left[
\wh\rho_{\wt\nu_1,\wt\nu_2}
\left(-\I\left(n+\frac12\right)\right)
-\frac{1}{(4\pi)^2}\log\mu_R^2
\right]
\operatorname*{Res}_{\wt\nu'=-\I(n+1/2)}
\wt{\mathcal I}^{(3)}_{\wt\nu'}(r).
\label{eq:loop-residue-sum}
\end{align}
The middle line of \eqref{eq:loop-residue-sum} vanishes at odd \(m\) because of the \(\sinh(\pi\wt\nu')\) factor in \eqref{eq:loop-antisymmetric-spectral-function}, and at even \(m\geq2\) because \(1/\Gamma(3/2+\I\wt\nu')=0\).  Only \(m=0\) remains, giving
\begin{align}
2\,\Pi_{\wt\nu_1,\wt\nu_2}\left(\frac{\I}{2}\right)
\operatorname*{Res}_{\wt\nu'=\I/2}
\wt{\mathcal I}^{(1)}_{\wt\nu'}(r)
={}&
-\frac{\wt\nu_1^2-\wt\nu_2^2}
{\cosh(2\pi\wt\nu_1)-\cosh(2\pi\wt\nu_2)}
\frac{r^3}{(1-r^2)^2}.
\label{eq:loop-surviving-tree-pole}
\end{align}

\subsection{Exact unequal-mass result and squeezed limit}

For coincident poles, including coincidences between spectral and folded-tree poles, the residue representation is understood as the continuous limit from generic masses.  Substituting \eqref{eq:loop-antisymmetric-spectral-function}, \eqref{eq:loop-spectral-poles}, \eqref{eq:loop-folded-signal-branch-1}, \eqref{eq:loop-folded-background}, and \eqref{eq:loop-surviving-tree-pole} into \eqref{eq:loop-residue-sum}, and using \eqref{eq:gamma-pole-residue} and \eqref{eq:dressed-hypergeometric}, we obtain:
\begin{align}\label{eq:loop-seed-unequal-mass}
\mathcal J_{\wt\nu_1,\wt\nu_2}(r)
=&~\bigg\{
\frac{1-\cosh\pi(\wt\nu_1+\wt\nu_2)}{16\pi^4}
\left(\frac{2r}{1+r}\right)^{4-\I\wt\nu_1-\I\wt\nu_2}
\sum_{n=0}^\infty \frac{(-1)^n}{n!}
\nonumber\\
&\times
\left(\frac{2r}{1+r}\right)^{2n}
\frac{3/2+2n-\I\wt\nu_1-\I\wt\nu_2}
{(1+2n-\I\wt\nu_1-\I\wt\nu_2)
 (2+2n-\I\wt\nu_1-\I\wt\nu_2)}
\nonumber\\
&\times \Gamma\left[
\frac32+n,\frac32+n-\I\wt\nu_1,
\frac32+n-\I\wt\nu_2,
\frac32+n-\I\wt\nu_1-\I\wt\nu_2
\right]
\nonumber\\
&\times \Gamma\left[
-n+\I\wt\nu_1,-n+\I\wt\nu_2,
-n+\I\wt\nu_1+\I\wt\nu_2
\right]
\nonumber\\
&\times {}_2\mathcal F_1\left[\begin{matrix}
2+2n-\I\wt\nu_1-\I\wt\nu_2,4+2n-\I\wt\nu_1-\I\wt\nu_2\\
4+4n-2\I\wt\nu_1-2\I\wt\nu_2
\end{matrix}\middle|\frac{2r}{1+r}\right]
\nonumber\\
&\quad
+(\wt\nu_1\to-\wt\nu_1)
+(\wt\nu_2\to-\wt\nu_2)
+\bigl((\wt\nu_1,\wt\nu_2)
\to(-\wt\nu_1,-\wt\nu_2)\bigr)
\bigg\}
\nonumber\\
&-  \frac{\wt\nu_1^2-\wt\nu_2^2}{\cosh 2\pi\wt\nu_1-\cosh 2\pi\wt\nu_2}\frac{r^3}{(1-r^2)^2}  \n\\
&+2\left(\frac{r}{1+r}\right)^3
\sum_{n=0}^\infty(-1)^n(2n+1)
\left[
\wh\rho_{\wt\nu_1,\wt\nu_2}
\left(-\I\left(n+\frac12\right)\right)
-\frac{1}{(4\pi)^2}\log\mu_R^2
\right]
\nonumber\\
&\hspace{1cm}\times
{}_3\mathcal F_2\left[
\begin{matrix}
1,1,3\\ 1-n,2+n
\end{matrix}\middle|\frac{2r}{1+r}\right].
\end{align}
Here and below, each of the three replacements acts on the original term in braces. The braces give the nonanalytic signal, while the final two terms are analytic at \(r=0\) and form the renormalized background.

In the squeezed limit, the leading background and nonanalytic terms are
\begin{align}\label{eq:loop-seed-unequal-mass-squeezed}
\lim_{r\to0}\mathcal J_{\wt\nu_1,\wt\nu_2}(r)
={}&
\bigg[ \frac{\I\tanh \frac\pi2 (\wt\nu_1+\wt\nu_2)}{\pi^2}\Big(\frac r2\Big)^{4-\I\wt\nu_1-\I\wt\nu_2}(3-\I\wt\nu_1-\I\wt\nu_2)\Gamma\Big[\frac32-\I\wt\nu_1,\frac32-\I\wt\nu_2, \I\wt\nu_1, \I\wt\nu_2 \Big] \n\\
&+(\wt\nu_1\to-\wt\nu_1)
+(\wt\nu_2\to-\wt\nu_2)
+\bigl((\wt\nu_1,\wt\nu_2)
\to(-\wt\nu_1,-\wt\nu_2)\bigr)
\bigg] \n\\
&+\left\{
4\left[
\wh\rho_{\wt\nu_1,\wt\nu_2}\left(-\frac{\I}{2}\right)
-\frac{1}{(4\pi)^2}\log\mu_R^2
\right]
-\frac{\wt\nu_1^2-\wt\nu_2^2}
{\cosh(2\pi\wt\nu_1)-\cosh(2\pi\wt\nu_2)}
\right\}r^3.
\end{align}

\subsection{Degenerate limit and physical continuation}

The unequal-mass results \eqref{eq:loop-seed-unequal-mass} and \eqref{eq:loop-seed-unequal-mass-squeezed} also admit the degenerate limit $\wt\nu_1=\wt\nu_2=\wt\nu$.  In this limit, we obtain
\begin{align}\label{eq:loop-seed-degenerate}
\mathcal J_{\wt\nu,\wt\nu}(r)
=&~\bigg\{
\frac{1-\cosh2\pi\wt\nu}{32\pi^4}
\left(\frac{2r}{1+r}\right)^{4-2\I\wt\nu}
\sum_{n=0}^\infty \frac{(-1)^n}{n!}
\nonumber\\
&\times
\left(\frac{2r}{1+r}\right)^{2n}
\frac{3/2+2n-2\I\wt\nu}
{(1+n-\I\wt\nu)(1+2n-2\I\wt\nu)}
\nonumber\\
&\times \Gamma\left[
\frac32+n,\frac32+n-\I\wt\nu,
\frac32+n-\I\wt\nu,\frac32+n-2\I\wt\nu
\right]
\nonumber\\
&\times \Gamma\left[
-n+\I\wt\nu,-n+\I\wt\nu,-n+2\I\wt\nu
\right]
\nonumber\\
&\times {}_2\mathcal F_1\left[\begin{matrix}
2+2n-2\I\wt\nu,4+2n-2\I\wt\nu\\
4+4n-4\I\wt\nu
\end{matrix}\middle|\frac{2r}{1+r}\right]
+(\wt\nu \to-\wt\nu)
\bigg\}
\nonumber\\
&-  \frac{\wt\nu\,\mathrm{csch}(2\pi\wt\nu)}{\pi}
\frac{r^3}{(1-r^2)^2}
\nonumber\\
&+ 2\Big(\frac{r}{1+r}\Big)^3
\sum_{n=0}^\infty(-1)^n (2n+1)
\left[
\wh\rho_{\wt\nu,\wt\nu}
\left(-\I\left(n+\frac12\right)\right)
-\frac{1}{(4\pi)^2}\log\mu_R^2
\right]
\nonumber\\
&\times {}_3\mathcal F_2\left[
\begin{matrix}
1,1,3\\ 1-n,2+n
\end{matrix}\middle|\frac{2r}{1+r}\right],
\end{align}
and its squeezed expansion is
\begin{align}\label{eq:loop-seed-degenerate-squeezed}
\lim_{r\to0}\mathcal J_{\wt\nu,\wt\nu}(r)
={}&
\frac{2\I\tanh \pi\wt\nu}{\pi^2}\Big(\frac r2\Big)^{4-2\I\wt\nu}\Gamma\Big[\frac32-\I\wt\nu,\frac52-\I\wt\nu, \I\wt\nu, \I\wt\nu \Big] + (\wt\nu\to-\wt\nu) \n\\
&+\left\{
4\left[
\wh\rho_{\wt\nu,\wt\nu}\left(-\frac{\I}{2}\right)
-\frac{1}{(4\pi)^2}\log\mu_R^2
\right]
-\frac{\wt\nu\,\mathrm{csch}(2\pi\wt\nu)}{\pi}
\right\}r^3.
\end{align}

For two identical complementary-series fields, we regard the complete renormalized seed as a meromorphic function of its mass parameter and continue \(\widetilde\nu\to-\I\nu\), retaining both mass-reflected branches before selecting the leading signal in \eqref{eq_J_Com_sq}. In the spectral representation, this continuation includes the contour deformation and the residues of any poles crossed along the way. At \(\nu=1/2\), the signal has the same \(r^3\) scaling as the analytic background, so their combined limit must be taken before separating them. At \(\nu=1\), the displayed leading signal coefficient vanishes.

\paragraph{Physical principal--complementary continuation}

Following \eqref{eq:propagator-mass-continuation}, for the mixed loop in Sec.~\ref{Loop} we apply the same mass convention as in App.~\ref{CalTree},
\begin{align}
\wt\nu_1&\longrightarrow-\I\nu_\pi,
&
\wt\nu_2&\longrightarrow\wt\nu_\sigma,
&
0<\nu_\pi&<\frac32,
&
\wt\nu_\sigma&>0.
\label{eq:loop-physical-continuation}
\end{align}
No spectral pole crosses the original contour in this range.  The continuation is therefore applied to the complete residue expressions, not to individual terms before the mass-reflected branches have been summed.  The leading nonanalytic powers in \eqref{eq:loop-seed-unequal-mass-squeezed} become \(r^{4-\nu_\pi\pm\I\wt\nu_\sigma}\), whereas the leading background is proportional to \(r^3\).  Consequently, the mixed signal decays more slowly than the background for \(1<\nu_\pi<3/2\).
\end{appendix}

\bibliographystyle{utphys}
\bibliography{references}

\end{document}